\documentclass[aps,preprint,amsmath,amssymb]{revtex4-1}
\usepackage{graphicx} 
\usepackage{amssymb,amsfonts,amsmath}
\usepackage{txfonts}
\usepackage[font=small]{caption} 
\usepackage{xcolor}              
\usepackage{setspace}            
\usepackage{fullpage}            
\usepackage{url}
\usepackage{subcaption}
\usepackage{booktabs}
\usepackage{url}
\usepackage{placeins}
\usepackage{epstopdf}

\definecolor{darkgreen}{HTML}{00BB00}

\begin{document}
\title{Robustness of a Network Formed of Spatially Embedded Networks}
\date{\today}
\author{Louis M. Shekhtman}
\author{Yehiel Berezin}
\author{Michael M. Danziger}
\author{Shlomo Havlin}
\affiliation{ Department of Physics, Bar-Ilan University, Ramat Gan, Israel}

\begin{abstract} 
 	 We present analytic and numeric results for percolation in a network formed of interdependent spatially embedded networks.
	We show results for a treelike and a random regular network of networks each with  $(i)$ unconstrained interdependent links and $(ii)$ interdependent links restricted to a maximum length, $r$. Analytic results are given for each network of networks with unconstrained dependency links and compared with simulations. For the case of two spatially embedded networks it was found that only for $r>r_c\approx8$ does the system undergo a first order phase transition.
	We find that  for treelike networks of networks $r_c$ significantly decreases as $n$ increases and rapidly reaches its limiting value, $r=1$. 
	For cases where the dependencies form loops, such as in random regular networks, we show analytically and confirm through simulations, that there is a certain fraction of dependent nodes, $q_{max}$, above which the entire network structure collapses even if a single node is removed. This $q_{max}$ decreases quickly with $m$, the degree of the random regular network of networks.  
	Our results show the extreme sensitivity of coupled spatial networks and emphasize the susceptibility of these networks to sudden collapse.
	The theory derived here can be used to find the robustness of any network of networks where the profile of percolation of a single network is known.
\\ \\
    \emph{Keywords:} mathematical and numerical analysis of networks, network stability under perturbation and duress, network percolation, interdependent networks
\end{abstract}
\maketitle

\section{Introduction}
As network science has expanded researchers have become aware of the fact that systems often consist of multiple interdependent networks \cite{rosato-criticalinf2008, parshani-prl2010, parshani-epl2010, shao-pre2011, leichtdsouza2009, cellai-pre2013, brummitt-pnas2012, xu-epl2011, hao2011interaction, PhysRevE.83.056208, PhysRevE.84.026101, huang-pre2011, gao-prl2011, hu-pre2011, bashan-pre2011, parshani-pnas2011, bashan-jsp2011, PhysRevE.85.066134, buldyrev-nature2010, gao-naturephysics2012, vespignani-nature2010, dong-preprint2012, bashan-naturephysics2013, rinaldi-ieee2001, peerenboom-proceedings2001, radicchi-naturephysics2013, son-epl2012, serrano-pre2012, morris-prl2012, zhao-jstatmech2013, donges-epjb2011, gao-general-net}. Examples of such systems are power grids that depend on communication networks, individuals who participate in multiple social circles, and metabolic networks that depend on other biological functions. Previous research on networks of networks provided a mathematical framework for understanding the stability of these systems \cite{gao-prl2011, PhysRevE.85.066134, gao-naturephysics2012,  leichtdsouza2009}. They found that these systems undergo a first-order percolation transition rather than the second-order transition which occurs for single networks. Recent work expanded the idea of interdependent networks to a pair of spatially embedded  networks \cite{wei-prl2012, bashan-naturephysics2013}. This represents an important step because many interdependent systems are spatially embedded \cite{brummitt-pnas2012, rinaldi-ieee2001, agarwal-milcom2010, wang-proceedings2008, albert-pre2004, kinney-epjb2005, chassin-physa2005, hines-chaos2010, carreras-ieee2004, barthelemy-physicsreports2011}. 

In our model, for each pair of connected networks a fraction $q_{ij}$ of nodes in network $i$ are assigned a dependent node in network $j$. The dependencies either follow the "no feedback condition", where if node $A$ in network $i$ depends on node $B$ in network $j$ then $B$ depends on $A$ as well or the "feedback condition" where such a constraint is not enforced \cite{gao-general-net}. For our simulations and theory we applied the "no feedback condition" and note that the "feedback condition" leads to an even more sensitive system. 

\begin{figure}
\centering

       \includegraphics[width=0.7\linewidth]{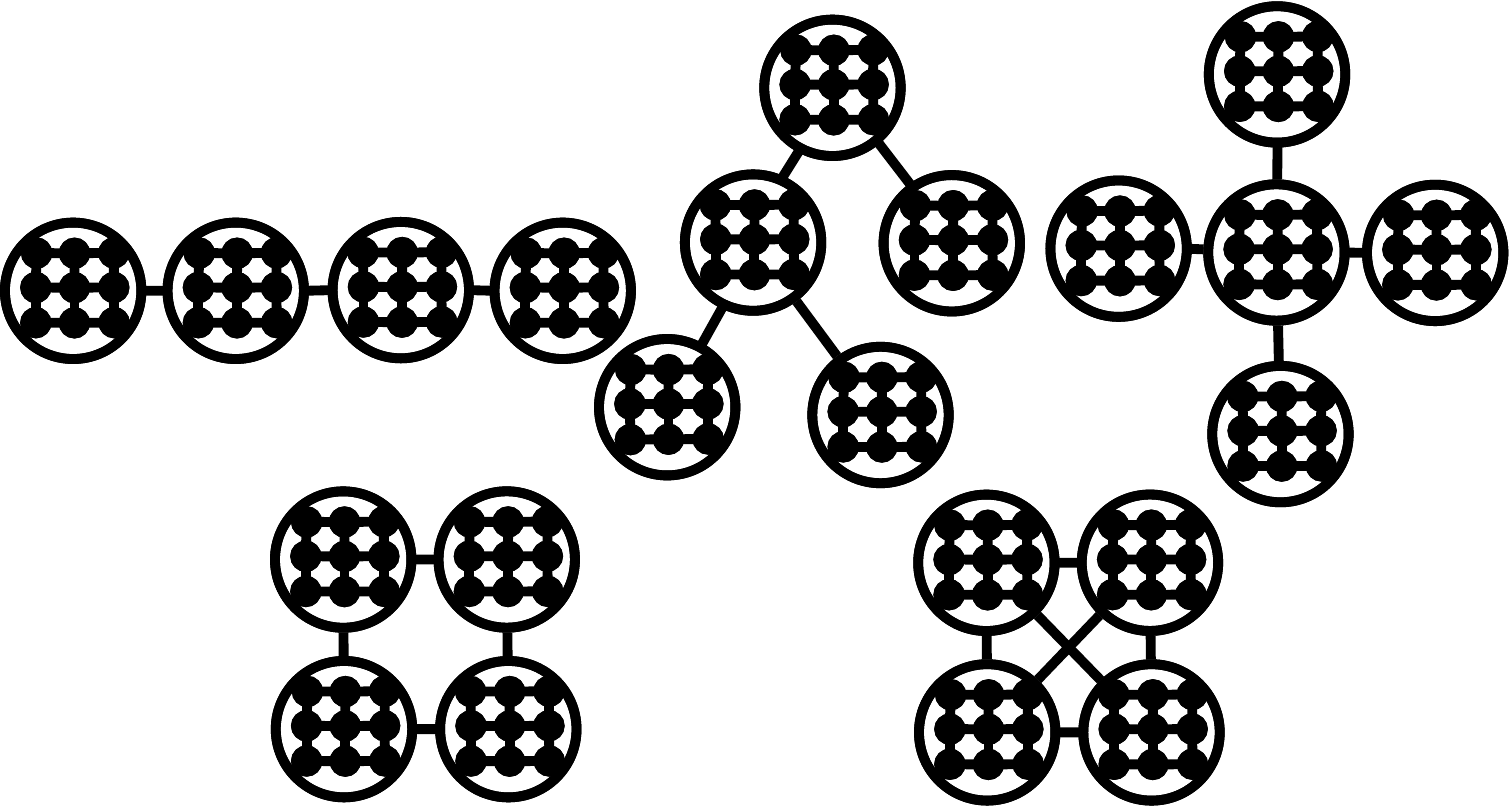}

\caption{ Several examples of possible structures of the network of networks. Examples include a line ({\bf top right}), a tree ({\bf top center}), a star  ({\bf top left}), a random regular network of networks where each network has $m=2$ dependencies  ({\bf bottom left}), and a random regular  network of networks with $m=3$  ({\bf bottom right}).}
	\label{fig:shapes}
\end{figure}

When $n > 2$ the dependencies can take on various topologies. We show some examples of possible configurations in Fig. ~\ref{fig:shapes}.

 At each step a fraction $1-p$ of the nodes are removed from either one or all of the networks. This leads to a dynamic cascade where each node removed causes dependent nodes to be removed in the other networks. 
The fraction of surviving nodes at the end of the cascade is defined as $x$. We then find $P_\infty(x)$, the mutual giant connected component where all remaining nodes are in their networks' respective giant component and where the dependencies of all nodes remaining are also still functional. 

 In order to simplify the study, we follow the method of previous studies \cite{wei-prl2012, bashan-naturephysics2013} where square lattices were used as the model of choice and note that any other 2D spatially embedded network with finite connectivity links belongs to the same universality class \cite{bunde1991fractals}. In the case of spatially embedded networks the dependencies are often restricted such that dependent nodes are within some distance, $r$, of one another \cite{wei-prl2012}. This quantity $r$ is called the dependency length and forces two dependent nodes $i$ and $j$, with positions $(x_i, y_i)$ and $(x_j, y_j)$ to obey $|x_i-x_j|\leq r$ and $|y_i-y_j|\leq r$ where $x$ and $y$ represent the positions of the nodes. The case $r=\infty$ is used to describe a network of networks with no restrictions on the maximum length of the dependency links. Previous work on interdependent spatially embedded networks involved only two networks and explored the interdependent fraction of the networks, $q$, and the maximum length of the dependency links, $r$ \cite{danziger_interdependent}. For each value of $q$ there is a critical dependency length, $r_c$, for which the percolation transition shifts from second order to first order. For $q=1$, i.e. two fully interdependent networks it was found $r_c\approx 8$ and for lower values of $q$ there is a higher value of $r_c$ \cite{danziger_interdependent}. It is of interest whether a higher number of interdependent systems could lead to a case where coupling between nearest or next nearest neighbors also leads to a first order transition.

 In line with previous research on networks of networks \cite{gao-general-net, gao-prl2011, PhysRevE.85.066134, gao-naturephysics2012} , we show simulation results for a network of networks with treelike dependencies and for a random regular network of networks, each with $(i)$ no restrictions on the length of dependency links and $(ii)$ dependency links of a maximum finite length, $r$. We also derive a theory for the size of the giant component of an interdependent system with no restrictions on the length of dependency links as a function of the system parameters and the percolation profile of a single lattice derived numerically. In our case this profile is obtained from the percolation profile of a $N=4000 \times 4000$ square lattice averaged over $100$ realizations. While we apply these equations to a lattice, we note that they can be used for any system composed of identical networks if $P_\infty(x)$ is known for the percolation of a single network.

\section{Interdependent Spatialy Embedded Networks with Tree-like Dependencies }
\subsection{Dynamics of cascading failures for a treelike network of networks}

We begin by examining results for a treelike network of networks where the length of dependency links is unconstrained. Li et~al. \cite{wei-prl2012} derived $P_\infty$ of the cascading failure of  two interdependent networks as a function of iteration count. It can be shown that if a fraction $1-p$ of nodes are removed from each network and $p_i$ is the fraction of survived nodes at the $i$th iteration $p_i=p^2\tfrac{P_\infty(p_{i-1})}{p_{i-1}}$. For $n$ networks in a treelike configuration a node is in the mutual giant component if  it and the $n-1$ nodes it depends on are all in their resepective networks' giant components. Thus $g(p_i)=P_\infty(p_{i-1})/p_{i-1}$, the probability for a node to be in the giant component after $1-p_i$ fraction of nodes are removed,  must be raised to the $n-1$ power since each node has $n-1$ dependencies. This gives
\begin{equation}
p_i=p^n\left(\tfrac{P_\infty(p_{i-1})}{p_{i-1}} \right)^{n-1}.
\label{eq:pi}
\end{equation}

Each iteration represents reducing all networks to their giant components and removing nodes which have dependencies outside the giant component. The next iteration then factors in the nodes removed due to having dependencies outside the previous giant component and again reduces each network to its giant component. The process repeats until a steady state is reached. In the limiting case of only a single network, $n=1$, we get $p_i=p$ and there is no cascading effect. Further if $n=2$, we obtain the known result \cite{wei-prl2012}.

\begin{figure}
\centering
\hfill
\begin{subfigure}{0.5\textwidth}
\centering
       \includegraphics[width=1.0\linewidth]{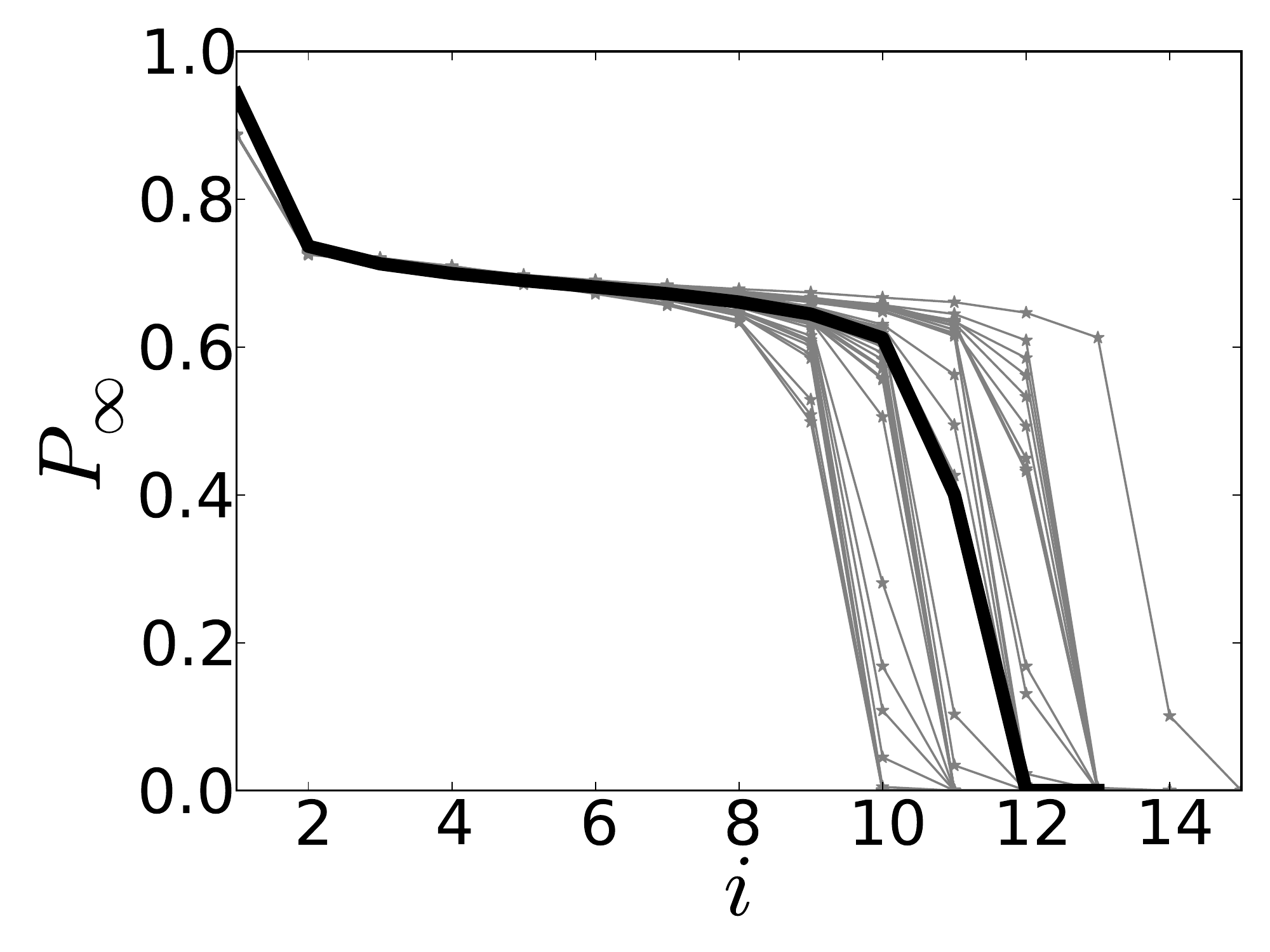}
	\caption{}
	\label{fig:pinf-iter}
\end{subfigure}%
\hfill
\begin{subfigure}{0.5\textwidth}
       \includegraphics[width=1.0\linewidth]{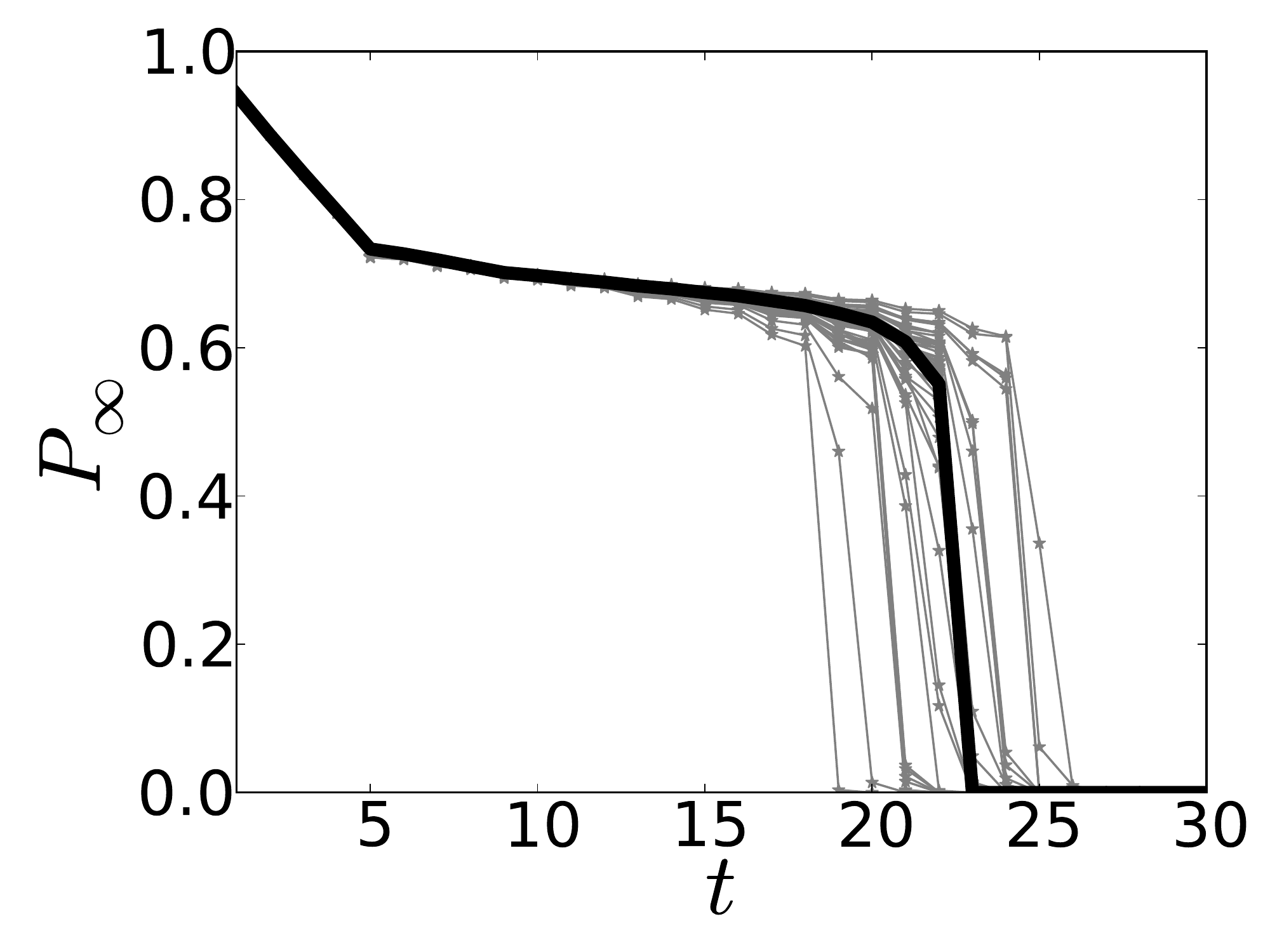}
	\caption{}
    \label{fig:pint-iter-gao}
\end{subfigure}
\hfill

\caption{Theory is shown as the thick black curve and 40 simulated realizations  on lattices of size $N=500\times500$ are shown as the lighter curves with symbols. {\bf (a) } $P_\infty$ as a function of the number of iterations is shown to fit well with the theory of Eq. (\ref{eq:pi}). These results are for five networks in a line, yet for this method the shape of the tree has no effect on the number of iterations.  {\bf (b)}  $P_\infty$ as a function of $t$ according to the method in  Gao et~al. \cite{PhysRevE.85.066134} for five networks in a line fits well with the theory.  }
\label{fig:time-iter}
\end{figure}

An alternate method of counting involves observing how the failures propogate across the links in the network of networks \cite{PhysRevE.85.066134}. The initial attack on each network occurs at $t=1$. A node which depends on a failed node then fails at $t=2$ and in general for a node that failed at $t=t_n$, its dependent nodes fail at $t=t_n+1$. Simulations  of the giant component after a number of iterations and at a certain time $t$ are shown to fit well with the theoretical equations in Fig. ~\ref{fig:time-iter}. 

\begin{figure}
\centering

       \includegraphics[width=0.7\linewidth]{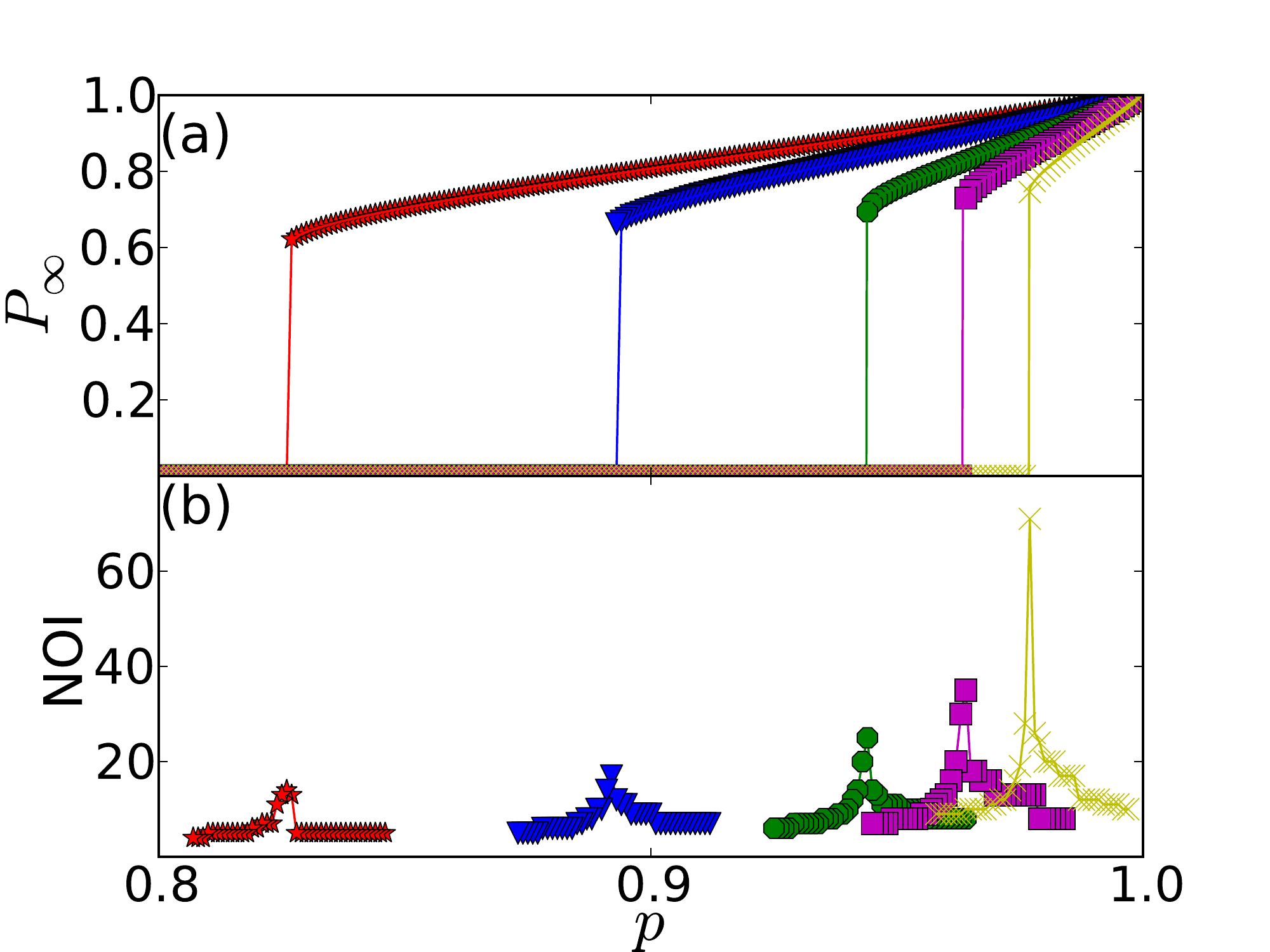}

\caption{{\bf (a)}  Both theory (lines) and simulations (symbols) for lattice networks of size $N=250 \times 250$  with treelike dependencies  are shown. These results are again for a network of networks in a line yet the results are the same for other tree formations. Results are shown for  $n=2$ (stars), $n=3$ (triangles), $n=5$ (circles), $n=7$ (squares), and $n=10$ ('x's). As seen all of the transitions are first order and the simulations fit well with the theory. Further, increasing the number of networks is seen to quickly increase $p_c$ indicating that the system becomes more sensitive as $n$ increases. {\bf (b)}  Here we observe that the number of iterations it takes for the system to arrive at steady-state diverges at $p_c$. The number of iterations at $p_c$ increases both with the number of networks, $n$, and the size of the networks, $N$ \cite{zhou2012critical}.  }
	\label{fig:rinf-tree-pinf}
\end{figure}

\subsection{Size of the giant component after the cascade}
We examine what happens to $p_i$ when $i\rightarrow \infty$ and the system reaches steady state. We define $x\equiv p_\infty$ and note that $x$ represents the total fraction of  nodes removed after the cascade including those removed due to interdependencies. For a given fraction of nodes, $1-p$, removed from the network of networks, $1-x$ is the fraction that would have to be removed from a single network to obtain an equivalent giant component.  Solving for $x$ we get
\begin{equation}
x=p\sqrt[n]{P_\infty(x)^{n-1}}.
\label{eq:pinf-tree-q1}
\end{equation}
In Fig. ~\ref{fig:rinf-tree-pinf} we observe that the theory agrees with simulations for all values of $p$. We also see there that close to $p_c$, the percolation threshold, the system collapses through a long cascade. The number of iterations at $p_c$ and $p_c$ both increase as the number of networks increases. 

To calculate $p_c$ we must find where the two sides of Eq. (\ref{eq:pinf-tree-q1}) are tangent at their intersection. We take the derivatives of both sides and get
\begin{align}
\tfrac{n}{n-1}P_\infty(x_c)&=x_cP'_\infty(x_c) \label{eq:tree-perc-all0} \\
p_c&=\tfrac{x_c}{P_\infty(x_c)^{(n-1)/n} \label{eq:tree-perc-all1}}
\end{align}
where $P'_\infty(x_c)$ is the derivative of $P_\infty(x_c)$. 

If a fraction $1-p$ is removed from only a single network, then $p^n$ in all the above equations is replaced with $p$. This gives 
\begin{align} 
x&=\sqrt[n]{pP_\infty(x)^{n-1}}. \label{eq:x-tree-all}\\ 
\tfrac{n}{n-1}P_\infty(x_c)&=x_cP'_\infty(x_c)   \label{eq:tree-pc-single0} \\ 
p_c&=\tfrac{x_c^{1/n}}{ P_\infty(x_c)^{(n-1)/n}}. 
\label{eq:tree-pc-single}
\end{align}
Results for $n$ networks according to Eqs. (\ref{eq:tree-pc-single0}) and (\ref{eq:tree-pc-single}) are shown in Fig. \ref{fig:star-pc} as the top curve ($q=1.0$). 
\subsection{ Effect of $q$, on $p_c$ for a starlike network of spatial networks}
\begin{figure}
\hfill
\begin{subfigure}{0.5\textwidth}
\centering
       \includegraphics[width=1.0\linewidth]{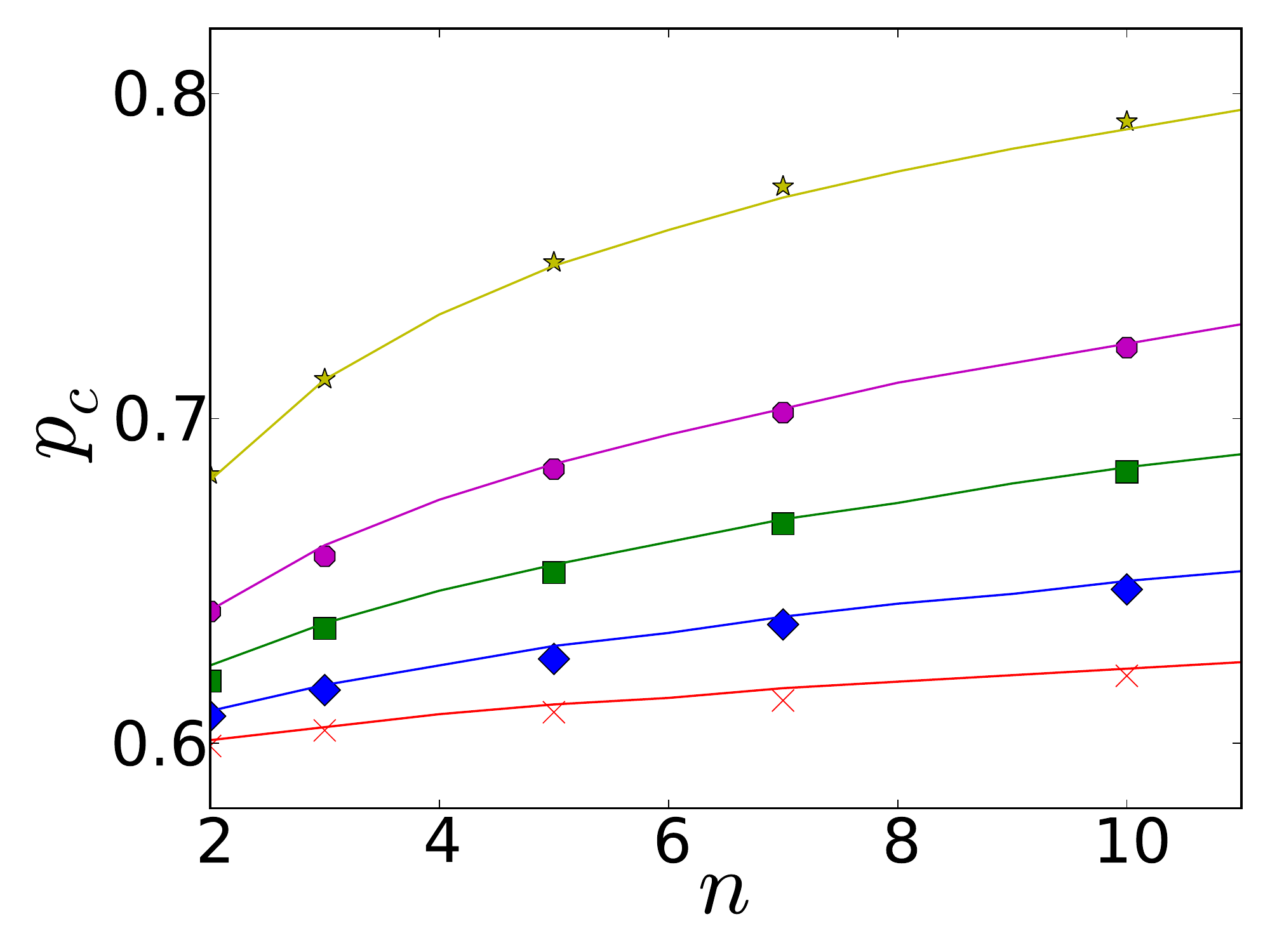}
	\caption{}
    \label{fig:rinf-tree-pc}
\end{subfigure}%
\hfill
\begin{subfigure}{0.5\textwidth}
\centering
       \includegraphics[width=1.0\linewidth]{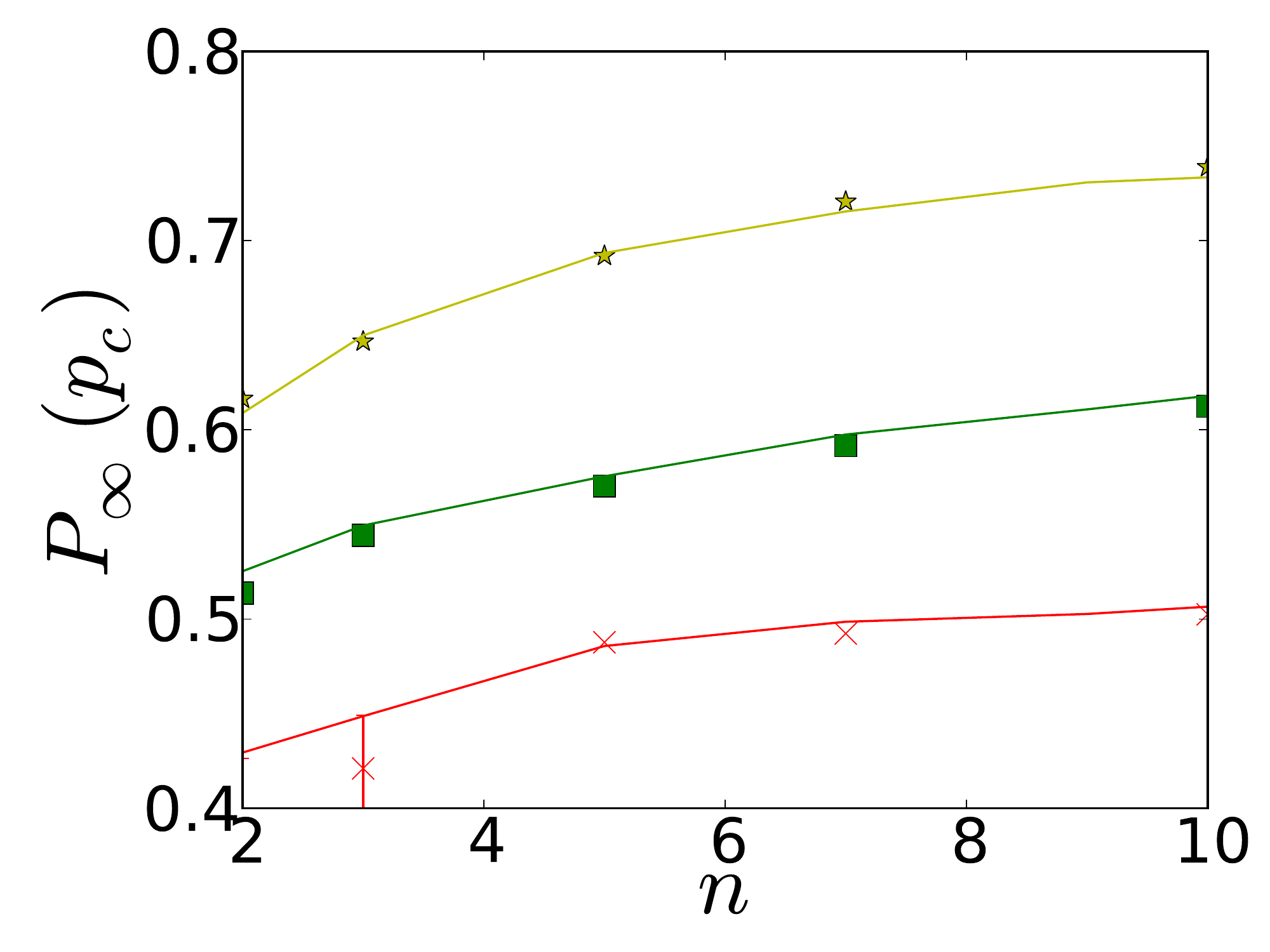}
	\caption{}
    \label{fig:rinf-tree-pinf-pc}
\end{subfigure}
\hfill

	\caption{{\bf a} The critical threshold $p_c$ as a function of the number of networks in a star formation is plotted as a function of $n$, the number of networks, for $q=0.5$ ('x's), $q=0.6$ (diamonds), $q=0.7$ (squares), $q=0.8$ (circles), and $q=1.0$ (stars). A fraction $1-p$ of nodes are removed only from the central network. Simulations on lattices with $N=250\times250$ ($n>2$) or $N=500\times500$  ($n=2$) show excellent agreement with the theory. {\bf b} $P_\infty(p_c)$ the size of the giant component at criticality is shown as a function of the number of networks (symbols are as before). Simulations on lattices of size $N=500\times500$ fit well with the theory. Where shown, errorbars represent a $1\sigma$ deviation.}
	\label{fig:star-pc}
\end{figure}

If we restrict the shape of the tree to be in a star formation with unrestricted dependency links (see Fig. ~\ref{fig:shapes}) we can also give an analytic solution for any value of the coupling $q$. In this case a fraction $1-p$ of the nodes are removed only from the central network. Based on Gao et~al. \cite{gao-prl2011} the equations for this system are
\begin{equation}
\begin{aligned}
x_1&=p(q\tfrac{P_{\infty_2}(x_2)}{x_2}-q+1)^{n-1} \\
x_2&=pq\tfrac{P_{\infty_1}(x_1)}{x_1}(q\tfrac{P_{\infty_2}(x_2)}{x_2}-q+1)^{n-2}-q+1 
\end{aligned}
\label{eq:star}
\end{equation}
where the subscript $1$ refers to the central network and the subscript $2$ refers to all the other networks. Results of theory and simulations for $p_c$ and $P_\infty(p_c)$ can be seen in Fig. ~\ref{fig:star-pc}. If $q=1$ Eq. (\ref{eq:star}) reduces to Eq. (\ref{eq:pinf-tree-q1}).

\subsection{Interdependent lattices with treelike dependencies with finite $r$}
\begin{figure}
\hfill
\begin{subfigure}{0.5\textwidth}
\centering
       \includegraphics[height=0.7\linewidth]{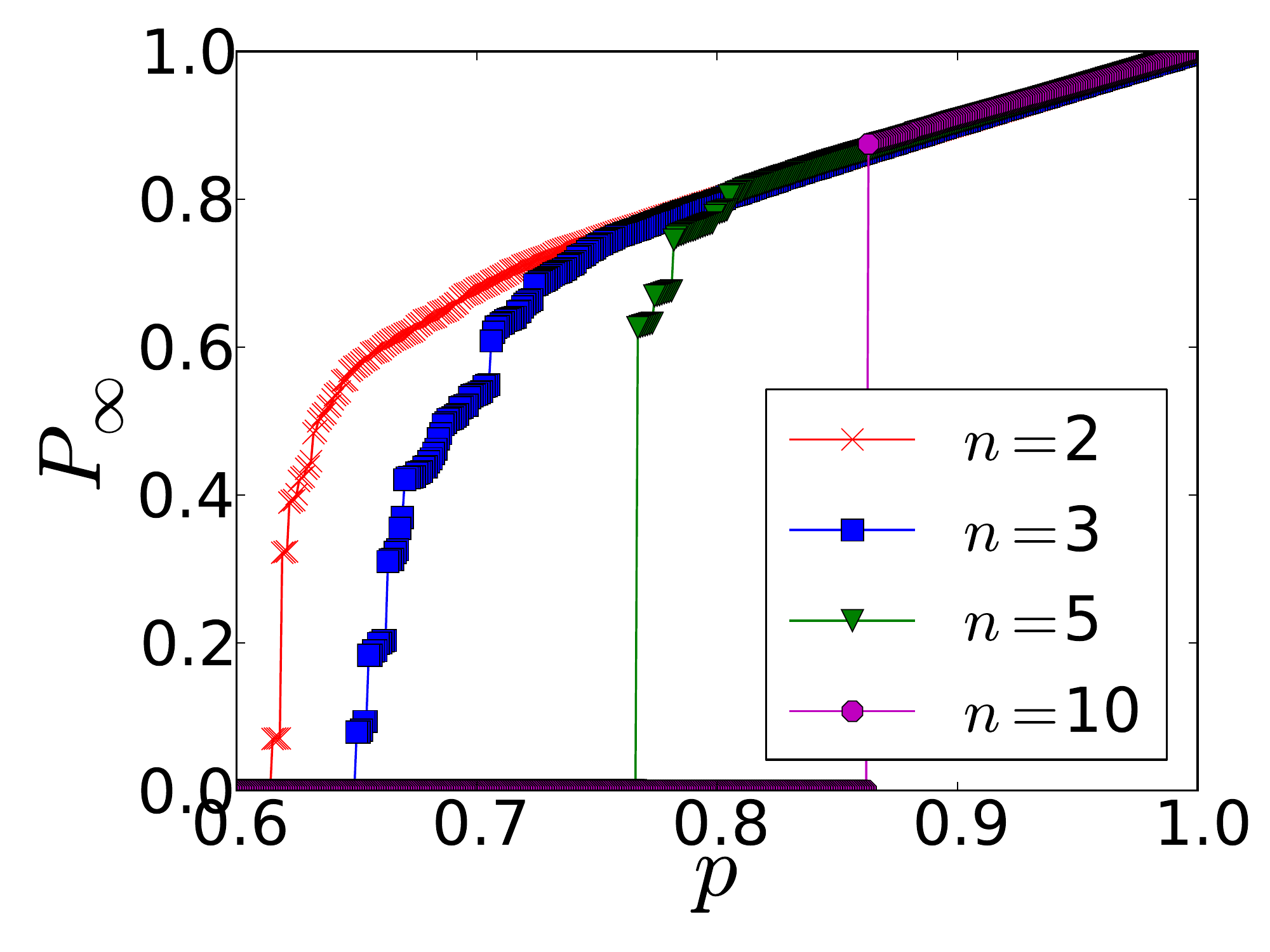}
\caption{}
    \label{fig:tree-finite-r}
\end{subfigure}%
\hfill
\begin{subfigure}{0.5\textwidth}
\centering
       \includegraphics[height=0.7\linewidth]{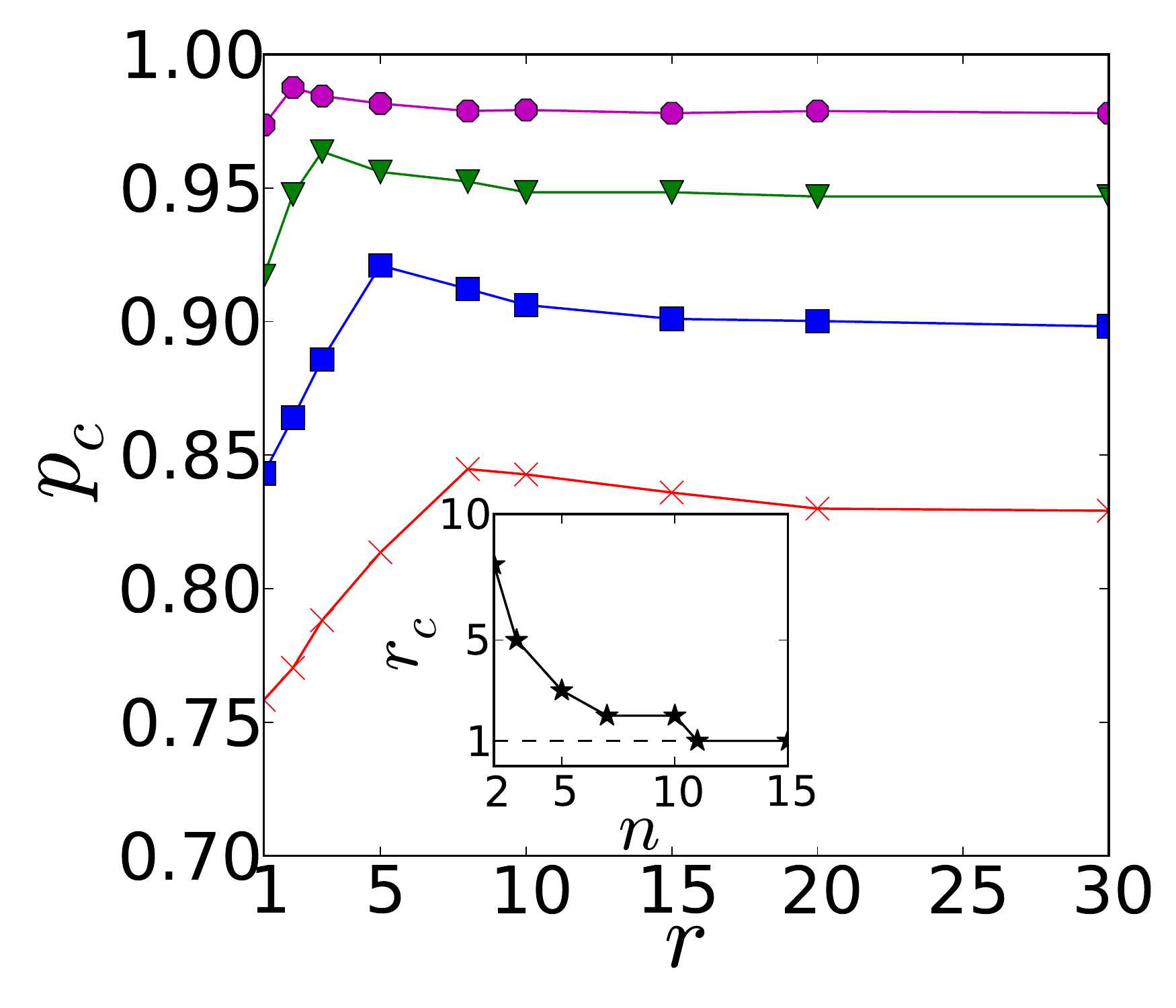}
\caption{ }
    \label{fig:tree-finite-r-pc}
\end{subfigure}
\hfill
\caption{{\bf (a)} Simulations for $n$ networks of size $N=100\times100$ in a tree formation with $r=2$ are shown. As the number of networks increases the transition becomes first-order. {\bf (b)} The percolation threshold $p_c$ is plotted as a function of $r$ for lattices with $N=250\times250$ with symbols as defined in Fig. ~\ref{fig:tree-finite-r}. The change from a second order to a first order transition occurs when $p_c$ reaches a maximum. The insert shows how this critical value, $r_c$, varies with the number of networks $n$. The critical dependency length can reach as low as $r_c=1$ if there is a sufficient number of interdependent networks. Note that already at $r=30$ the results begin to agree with the theory of Eqs. (\ref{eq:tree-perc-all0}) and (\ref{eq:tree-perc-all1}) for $r=\infty$. In this case since a fraction $1-p$ was removed from all networks, we must take $p_c^{1/n}$ in order to get results that agree with Eqs. (\ref{eq:tree-pc-single0}) and (\ref{eq:tree-pc-single}) and Fig. \ref{fig:rinf-tree-pc}. }
\end{figure}

We now examine a network of networks with treelike dependencies but now with a finite maximum dependency length, i.e. $r<\infty$. Due to spatial constrains, theory becomes a difficult task and we limit ourselves to the simulations. We remove a fraction $1-p$ of the nodes from each network and note that the results can be converted to a case where  nodes are removed from just a single network using $p^n \rightarrow p$. Previous research on a pair of interdependent lattices found that $p_c$ displays rich behavior as $r$ is increased \cite{wei-prl2012, danziger_interdependent}. When $r$ is small, $p_c$ increases linearly with $r$ until it reaches a peak. This peak represents the point where the system changes from a second order transition to a first order transition. As $r$ increases past the peak $p_c$ decreases and approaches its limiting value at $r=\infty$. 
 
First we analyze the giant component as a function of $p$ for different numbers of networks in a tree. In Fig. ~\ref{fig:tree-finite-r} we observe that the system now undergoes a first order transition even for $r=2(<8)$ if there are a sufficient number of networks. Our simulations reveal that the results are the same (for $q=1$) regardless of whether the network of networks is in a star or a line, i.e. the results are independent of the shape of the tree. From the peak of the graph $p_c$ vs. $r$ in Fig. ~\ref{fig:tree-finite-r-pc}, we are able to determine $r_c$, the critical dependency length where the collapse becomes first order. In the inset of Fig. ~\ref{fig:tree-finite-r-pc} we observe that this critical dependency length decreases significantly as we increase the number of networks in the tree.   When $n=11$, we get $r_c=1$ which is its limiting value.

\section{Interdependent Lattices with Random Regular Dependencies}

\subsection{Random regular network of spatial networks with random dependencies}
\begin{figure}
\centering
\hfill
\begin{subfigure}{.5\textwidth}
	\centering
       \includegraphics[width=1.0\linewidth]{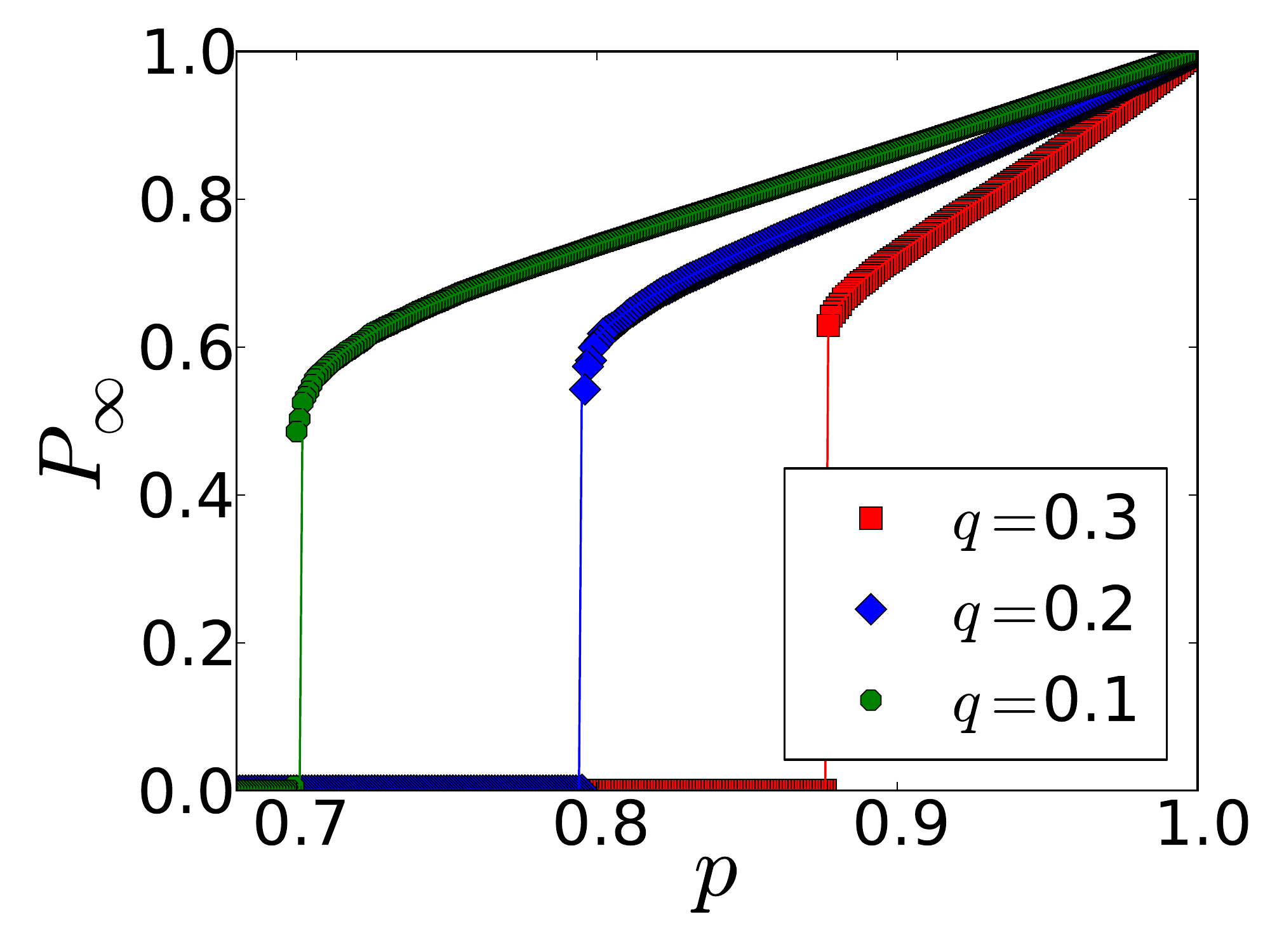}
	\caption{}
    \label{fig:rinf-RR-pinf-qvals}

\end{subfigure}%
\hfill
\begin{subfigure}{.5\textwidth}
\centering
       \includegraphics[width=1.0\linewidth]{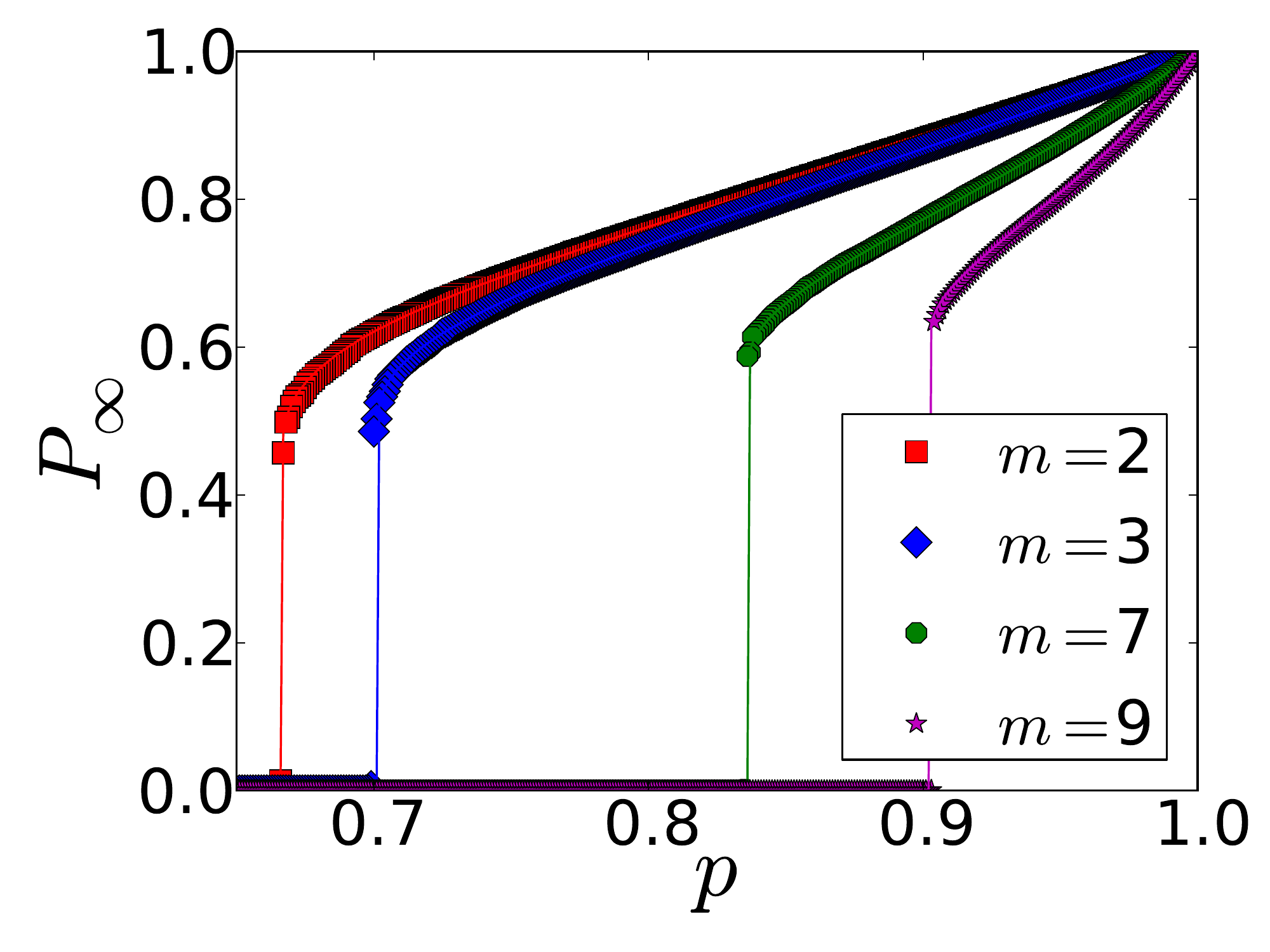}
	\caption{}
    \label{fig:rinf-RR-pinf-mvals}

\end{subfigure}
\hfill

\caption{ The giant components according to both theory and simulations for interdependent lattice networks of size $N=250\times250$ with random regular dependencies where $r=\infty$ are shown. As seen all of the transitions are first order and the simulations fit well with the theory. We use $n=m+1$ networks in the simulations, but the results depend only on $m$, the number of dependencies, each network has and not on $n$, the actual number of networks in the system.  Results for {\bf (a) } different values of $q$ and  {\bf (b) } different values of $m$ are shown. }
\label{fig:rinf-RR-pinf}
\end{figure}

Many real networks contain loops and thus we now derive results for a random regular network of spatial networks. Gao et al. \cite{gao-naturephysics2012} previously showed that for such a dependency configuration the actual number of networks, $n$ is irrelevant, rather the results depend only on $m$ the number of networks each network depends on. In this case if all nodes are interdependent ($q=1$) the network collapses immediately due to the loops. We therefore limit our analysis to $q<1$ and remove a fraction $1-p$ of the nodes from all the networks. Gao et al. \cite{gao-naturephysics2012} obtained
\begin{equation}
\begin{cases}
x=p(qyg(x)-q+1)^m, \\
y=p(qyg(x)-q+1)^{m-1}
\end{cases}
\label{eq:RR-pinf0}
\end{equation}
where $y$ represents the percolation damage from all networks except from the dependency link currently being examined. The system in Eq. (\ref{eq:RR-pinf0}) can be solved by eliminating $y$ from the second equation and obtaining a single equation for $x$. After substituting $g(x)=P_\infty(x)/x$ we obtain

\begin{equation}
P_\infty(x)p^{2/m}q=x^{2/m}+(xp)^{1/m}(q-1)
\label{eq:RR-Pinf}
\end{equation}
which can be solved numerically for $x$ given any values of $p$, $q$ and $m$. Simulations and theory according to Eq. (\ref{eq:RR-Pinf}), based on the numerical form of $P_\infty(x)$ for a single lattice, are shown in Fig. ~\ref{fig:rinf-RR-pinf}.

To derive $p_c$ we take derivatives of both sides of Eq. (\ref{eq:RR-Pinf}) and obtain
\begin{equation}
mx_cP'_\infty(x_c)\left( q-1\pm\sqrt{(q-1)^2+4qP_\infty(x_c)} \right)^2=
8qP_\infty(x_c)^2+2(q-1)P_\infty(x_c) \left(q-1\pm \sqrt{(q-1)^2+4qP_\infty(x_c)}\right).
\label{eq-RR-xc}
\end{equation}
which can be solved for $x_c$. From Eq. (\ref{eq:RR-Pinf}) we can get $p_c$  by rearranging Eq. (\ref{eq:RR-Pinf}) to
\begin{equation}
p_c=\left[\tfrac{x_c^{1/m}}{2qP_\infty(x_c)}\left( q-1\pm\sqrt{(q-1)^2+4qP_\infty(x_c)}\right)\right]^m.
\label{eq:RR-pc-final}
\end{equation}
Simulations and theory according to Eq. (\ref{eq:RR-pc-final}) are shown in Fig. ~\ref{fig:rinf-RR-pc}. 
\subsection{Maximum coupling, $q_{max}$}

\begin{figure}
\centering
\hfill
\begin{subfigure}{0.5\textwidth}
\centering
       \includegraphics[width=1.0\linewidth]{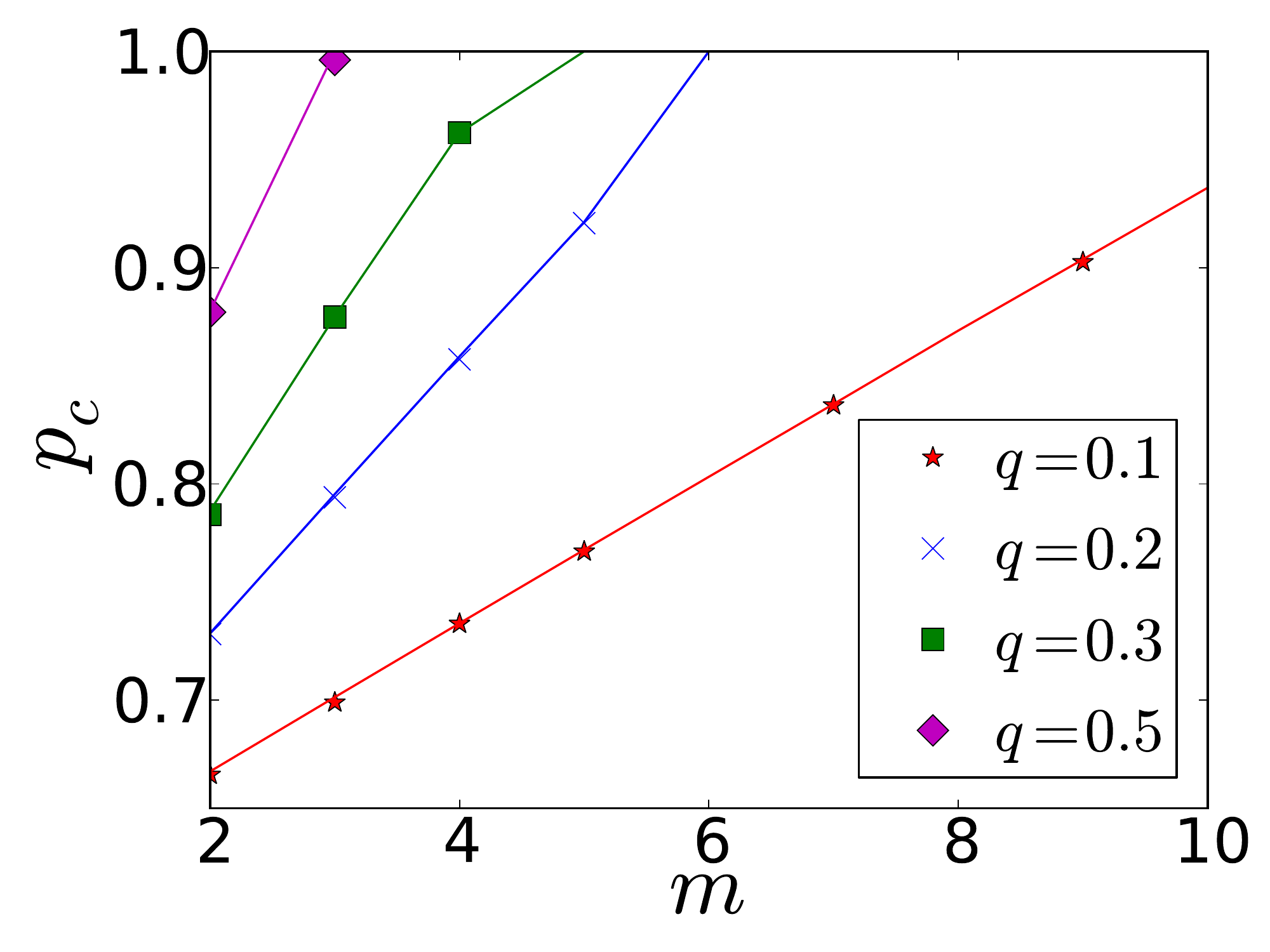}
\caption{}
    \label{fig:rinf-RR-pc}
\end{subfigure}%
\hfill
\begin{subfigure}{0.5\textwidth}
\centering
       \includegraphics[width=1.0\linewidth]{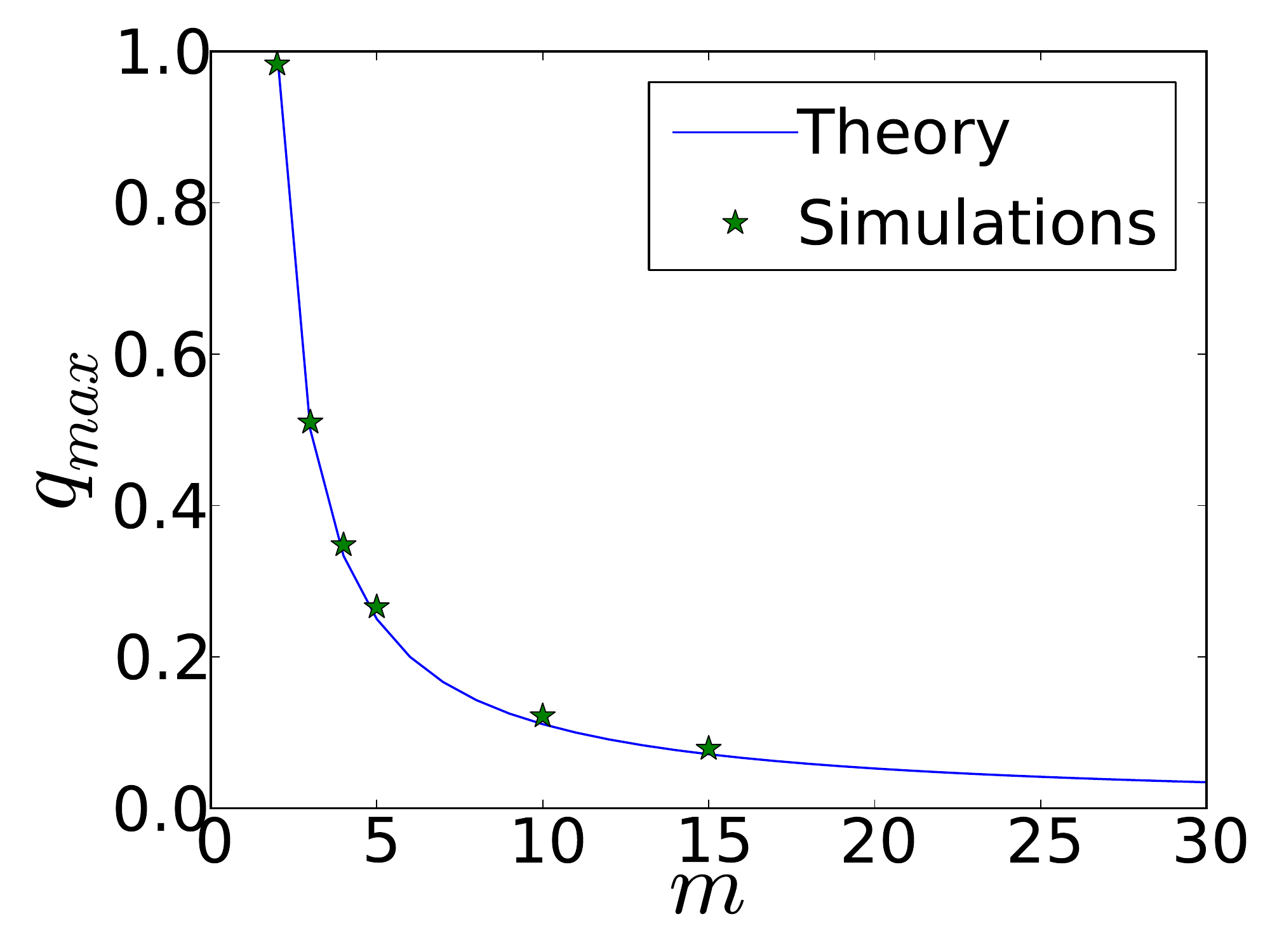}
\caption{}
    \label{fig:qmax-RR}
\end{subfigure}
\hfill
\caption{{\bf (a)}  The critical threshold $p_c$ is plotted as a function of $m$, the number of dependencies each network has for several values of $q$. The lines represent the theory according to Eq. (\ref{eq:RR-pc-final}) and the symbols represent simulations. It is worth noting that once the number of dependencies reaches a certain value, $p_c\rightarrow 1$ for a given $q$-value. {\bf (b)}   The  maximum coupling between networks, $q_{max}$, above which $p_c\rightarrow 1$ is plotted as a function of $m$. As seen, $q_{max}$ decreases quickly with $m$, indicating that as each network has more dependencies less coupling is required for the network to fail after even the smallest attack. Simulations (symbols) on lattices of size $N=250\times250$ are shown to fit well with the theory (line).}
\end{figure}

From the graph of $p_c$ in Fig. ~\ref{fig:rinf-RR-pc} it is clear that for each value of $m$ there is some maximum coupling $q_{max}$ for which removing even a single node will lead the entire network to collapse. We can solve for $q_{max}$ by using Eq. (\ref{eq:RR-Pinf}) and setting $p=1$. This gives 
\begin{equation}
q_{max}=\tfrac{x_{max}^{2/m}-x_{max}^{1/m}}{P_\infty(x_{max})-x_{max}^{1/m}}.
\label{eq:qmax}
\end{equation}

 We can then solve for $x_{max}$ using Eq. (\ref{eq-RR-xc}). Explicitly,

\begin{equation}
mx_{max}P'_\infty(x_{max})(x_{max}^{2/m}-x_{max}^{1/m})=-x_{max}^{3/m}+P_\infty(x_{max})(2x_{max}^{2/m}-x_{max}^{1/m}). 
\end{equation}
After we have $x_{max}$ we substitute it into Eq. (\ref{eq:qmax}) and obtain $q_{max}$, which is plotted in Fig. ~\ref{fig:qmax-RR}.

\subsection{Interdependent lattices with random regular dependencies with finite $r$}

\begin{figure}
\centering
\hfill
\begin{subfigure}{0.5\textwidth}
\centering
       \includegraphics[width=1.0\linewidth]{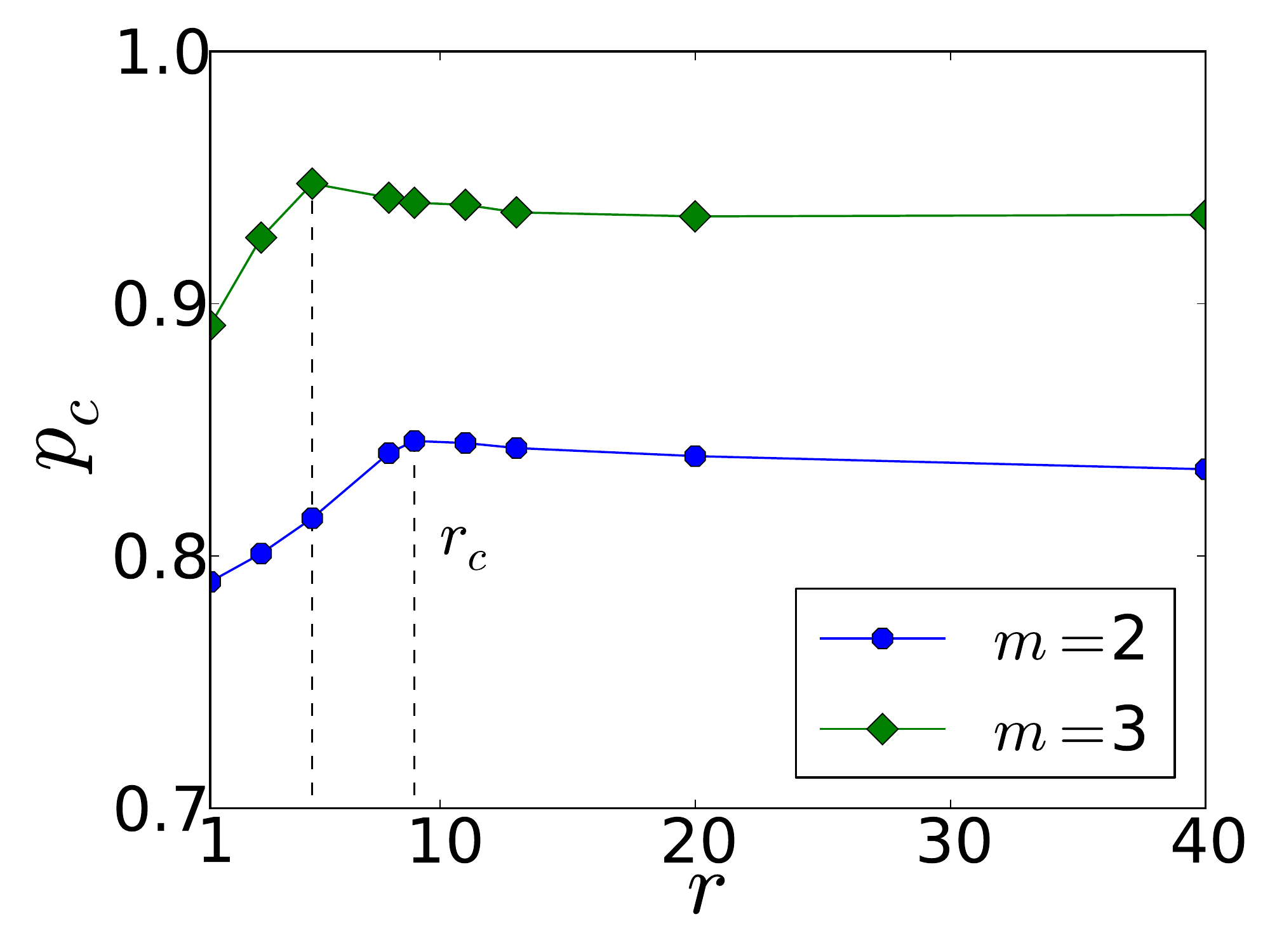}
	\caption{}
\end{subfigure}%
\hfill
\begin{subfigure}{0.5\textwidth}
\centering
       \includegraphics[width=1.0\linewidth]{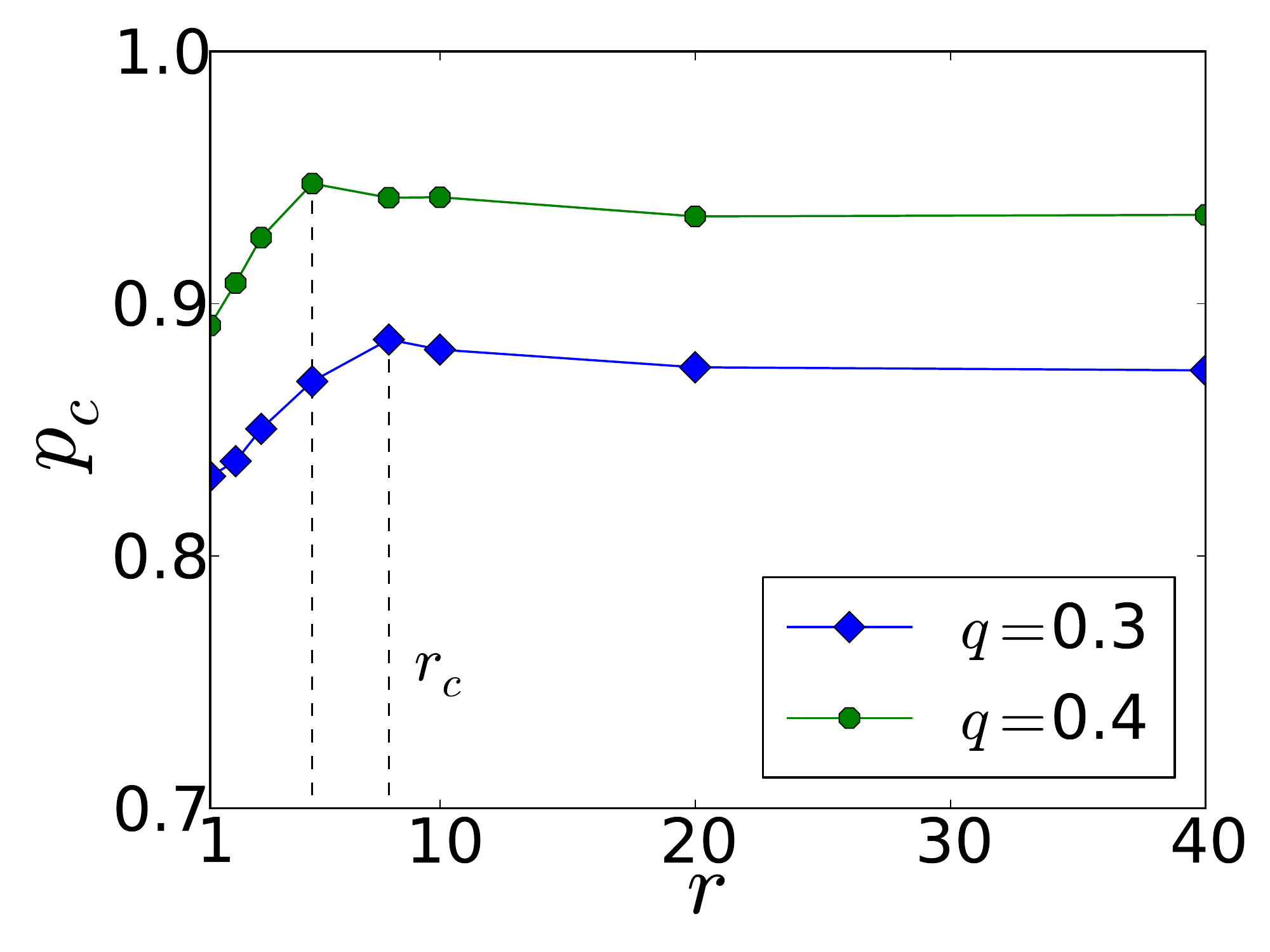}
	\caption{}
\end{subfigure}
\hfill
\caption{The shift from second order to first order transition occurs where $p_c$ reaches a maximum. This maximum is seen to occur for smaller $r$  as: {\bf (a)} the number of dependent networks increases (with $q=0.4$)  and {\bf (b)} the interdependent fraction, $q$, between the networks increases (with $m=3$). Simulations are performed on lattices of size $N=250\times250$.}
    \label{fig:RR-finite-r-pc}
\end{figure}

We now analyze random regular networks where dependency links are of a finite maximum length, $r$. We observe in Fig. ~\ref{fig:RR-finite-r-pc} that the shape of the $p_c$ vs. $r$ curve is the same as it was for trees and the transition switches from second order to first order above $r_c$ which is the value of $r$ when $p_c$ is at a maximum. As the number of neighbors, $m$, increases the critical dependency length $r_c$, decreases significantly for any value of $q$. Further, for high values of $q$ the  system collapses even for $m=2$ and $r=1$. Thus a system with as few as $3$ networks, fully interconnected ($m=2$), is extremely sensitive if there is a high level of interdependency. In Fig. ~\ref{fig:RR-finite-r-pc} we observe that $r_c$ decreases as $m$ increases and also as $q$ increases. 

Continuing the analysis that was done in Chap. {\bf  III}, section {\bf B}, for $r=\infty$, we explore for finite $r$, the values of $q_{max}$, the maximum coupling above which the system collapses. As was the case for infinite $r$, we find that $q_{max}$ for systems with finite $r$ drops rapidly as $m$ is increased. Further the difference for a given $m$ between $q_{max}$ with finite $r$ and $q_{max}$ with $r=\infty$ decreases as $m$ increases. For low values of $m$ introducing a finite $r$ leads to an increase in $q_{max}$  yet for $m \gtrsim 15$ there is almost no difference in $q_{max}$ for $r=\infty$ and $q_{max}$ for $r=1$.  Thus networks of spatial networks with many dependencies are extremely sensitive even for small $q$ and small $r$. 
\section{Discussion}
In summary, we have applied the framework for a network of networks to the case of $n$ spatially embedded networks. We provide analytic results for dependencies in a tree and a random regular configuration with no restrictions on the length of the dependency links. Further we provide simulations for these two cases and find excellent agreement with the theory. We also studied simulations for each case when dependency links are of a finite length, $r$. For $n$ networks in a tree configuration we find that the critical dependency length for a first order transition decreases significantly as more networks are added and with enough interdependent networks the system undergoes a first order transition even for $r=1$,  i.e. nearest neighbor dependencies.

If the dependencies contain loops we find that there is a critical value of coupling $q_{max}$ above which the system will collapse even if a single node is removed. This $q_{max}$ decreases significantly as the number of dependencies of each network increases and for high values of $m$ this $q_{max}$ is virtually unaffected by imposing a finite length on the dependency links. 

These results emphasize the vulnerability of interdependent spatially embedded networks and show that many colocalized interacting systems can collapse suddenly. Our model here can help explain sudden failures seen in many real-world systems such as powergrids.

\section*{Acknowledgments}
We acknowledge the European EPIWORK and MULTIPLEX (EU-FET project 317532)
projects, the Deutsche Forschungsgemeinschaft (DFG) and the Israel Science Foundation,
for providing ffinancial support.

\bibliographystyle{naturemag}
\bibliography{paper}

\begin{thebibliography}{10}
\expandafter\ifx\csname url\endcsname\relax
  \def\url#1{\texttt{#1}}\fi
\expandafter\ifx\csname urlprefix\endcsname\relax\def\urlprefix{URL }\fi
\providecommand{\bibinfo}[2]{#2}
\providecommand{\eprint}[2][]{\url{#2}}

\bibitem{rosato-criticalinf2008}
\bibinfo{author}{Rosato, V.} \emph{et~al.}
\newblock \bibinfo{title}{{Modelling interdependent infrastructures using
  interacting dynamical models}}.
\newblock \emph{\bibinfo{journal}{International Journal of Critical
  Infrastructures}} \textbf{\bibinfo{volume}{4}}, \bibinfo{pages}{63}
  (\bibinfo{year}{2008}).
\newblock \urlprefix\url{http://dx.doi.org/10.1504/IJCIS.2008.016092}.

\bibitem{parshani-prl2010}
\bibinfo{author}{Parshani, R.}, \bibinfo{author}{Buldyrev, S.~V.} \&
  \bibinfo{author}{Havlin, S.}
\newblock \bibinfo{title}{{Interdependent Networks: Reducing the Coupling
  Strength Leads to a Change from a First to Second Order Percolation
  Transition}}.
\newblock \emph{\bibinfo{journal}{Phys. Rev. Lett.}}
  \textbf{\bibinfo{volume}{105}}, \bibinfo{pages}{048701}
  (\bibinfo{year}{2010}).
\newblock
  \urlprefix\url{http://link.aps.org/doi/10.1103/PhysRevLett.105.048701}.

\bibitem{parshani-epl2010}
\bibinfo{author}{Parshani, R.}, \bibinfo{author}{Rozenblat, C.},
  \bibinfo{author}{Ietri, D.}, \bibinfo{author}{Ducruet, C.} \&
  \bibinfo{author}{Havlin, S.}
\newblock \bibinfo{title}{{Inter-similarity between coupled networks}}.
\newblock \emph{\bibinfo{journal}{EPL (Europhysics Letters)}}
  \textbf{\bibinfo{volume}{92}}, \bibinfo{pages}{68002} (\bibinfo{year}{2010}).
\newblock \urlprefix\url{http://stacks.iop.org/0295-5075/92/i=6/a=68002}.

\bibitem{shao-pre2011}
\bibinfo{author}{Shao, J.}, \bibinfo{author}{Buldyrev, S.~V.},
  \bibinfo{author}{Havlin, S.} \& \bibinfo{author}{Stanley, H.~E.}
\newblock \bibinfo{title}{{Cascade of failures in coupled network systems with
  multiple support-dependence relations}}.
\newblock \emph{\bibinfo{journal}{Phys. Rev. E}} \textbf{\bibinfo{volume}{83}},
  \bibinfo{pages}{036116} (\bibinfo{year}{2011}).
\newblock \urlprefix\url{http://link.aps.org/doi/10.1103/PhysRevE.83.036116}.

\bibitem{leichtdsouza2009}
\bibinfo{author}{{Leicht}, E.~A.} \& \bibinfo{author}{{D'Souza}, R.~M.}
\newblock \bibinfo{title}{{Percolation on interacting networks}}.
\newblock \emph{\bibinfo{journal}{ArXiv e-prints}}  (\bibinfo{year}{2009}).
\newblock \eprint{0907.0894}.

\bibitem{cellai-pre2013}
\bibinfo{author}{Cellai, D.}, \bibinfo{author}{L\'opez, E.},
  \bibinfo{author}{Zhou, J.}, \bibinfo{author}{Gleeson, J.~P.} \&
  \bibinfo{author}{Bianconi, G.}
\newblock \bibinfo{title}{Percolation in multiplex networks with overlap}.
\newblock \emph{\bibinfo{journal}{Phys. Rev. E}} \textbf{\bibinfo{volume}{88}},
  \bibinfo{pages}{052811} (\bibinfo{year}{2013}).
\newblock \urlprefix\url{http://link.aps.org/doi/10.1103/PhysRevE.88.052811}.

\bibitem{brummitt-pnas2012}
\bibinfo{author}{Brummitt, C.~D.}, \bibinfo{author}{D'Souza, R.~M.} \&
  \bibinfo{author}{Leicht, E.~A.}
\newblock \bibinfo{title}{{Suppressing cascades of load in interdependent
  networks}}.
\newblock \emph{\bibinfo{journal}{Proceedings of the National Academy of
  Sciences}} \textbf{\bibinfo{volume}{109}}, \bibinfo{pages}{E680--E689}
  (\bibinfo{year}{2012}).

\bibitem{xu-epl2011}
\bibinfo{author}{Xu, X.-L.}, \bibinfo{author}{Qu, Y.-Q.},
  \bibinfo{author}{Guan, S.}, \bibinfo{author}{Jiang, Y.-M.} \&
  \bibinfo{author}{He, D.-R.}
\newblock \bibinfo{title}{{Interconnecting bilayer networks}}.
\newblock \emph{\bibinfo{journal}{EPL (Europhysics Letters)}}
  \textbf{\bibinfo{volume}{93}}, \bibinfo{pages}{68002} (\bibinfo{year}{2011}).
\newblock \urlprefix\url{http://stacks.iop.org/0295-5075/93/i=6/a=68002}.

\bibitem{hao2011interaction}
\bibinfo{author}{Hao, J.}, \bibinfo{author}{Cai, S.}, \bibinfo{author}{He, Q.}
  \& \bibinfo{author}{Liu, Z.}
\newblock \bibinfo{title}{{The interaction between multiplex community
  networks}}.
\newblock \emph{\bibinfo{journal}{Chaos: An Interdisciplinary Journal of
  Nonlinear Science}} \textbf{\bibinfo{volume}{21}},
  \bibinfo{pages}{016104--016104} (\bibinfo{year}{2011}).

\bibitem{PhysRevE.83.056208}
\bibinfo{author}{Morino, K.}, \bibinfo{author}{Tanaka, G.} \&
  \bibinfo{author}{Aihara, K.}
\newblock \bibinfo{title}{{Robustness of multilayer oscillator networks}}.
\newblock \emph{\bibinfo{journal}{Phys. Rev. E}} \textbf{\bibinfo{volume}{83}},
  \bibinfo{pages}{056208} (\bibinfo{year}{2011}).
\newblock \urlprefix\url{http://link.aps.org/doi/10.1103/PhysRevE.83.056208}.

\bibitem{PhysRevE.84.026101}
\bibinfo{author}{Gu, C.-G.} \emph{et~al.}
\newblock \bibinfo{title}{{Onset of cooperation between layered networks}}.
\newblock \emph{\bibinfo{journal}{Phys. Rev. E}} \textbf{\bibinfo{volume}{84}},
  \bibinfo{pages}{026101} (\bibinfo{year}{2011}).
\newblock \urlprefix\url{http://link.aps.org/doi/10.1103/PhysRevE.84.026101}.

\bibitem{huang-pre2011}
\bibinfo{author}{Huang, X.}, \bibinfo{author}{Gao, J.},
  \bibinfo{author}{Buldyrev, S.~V.}, \bibinfo{author}{Havlin, S.} \&
  \bibinfo{author}{Stanley, H.~E.}
\newblock \bibinfo{title}{{Robustness of interdependent networks under targeted
  attack}}.
\newblock \emph{\bibinfo{journal}{Phys. Rev. E}} \textbf{\bibinfo{volume}{83}},
  \bibinfo{pages}{065101} (\bibinfo{year}{2011}).
\newblock \urlprefix\url{http://link.aps.org/doi/10.1103/PhysRevE.83.065101}.

\bibitem{gao-prl2011}
\bibinfo{author}{Gao, J.}, \bibinfo{author}{Buldyrev, S.~V.},
  \bibinfo{author}{Havlin, S.} \& \bibinfo{author}{Stanley, H.~E.}
\newblock \bibinfo{title}{{Robustness of a Network of Networks}}.
\newblock \emph{\bibinfo{journal}{Phys. Rev. Lett.}}
  \textbf{\bibinfo{volume}{107}}, \bibinfo{pages}{195701}
  (\bibinfo{year}{2011}).
\newblock
  \urlprefix\url{http://link.aps.org/doi/10.1103/PhysRevLett.107.195701}.

\bibitem{hu-pre2011}
\bibinfo{author}{Hu, Y.}, \bibinfo{author}{Ksherim, B.},
  \bibinfo{author}{Cohen, R.} \& \bibinfo{author}{Havlin, S.}
\newblock \bibinfo{title}{{Percolation in interdependent and interconnected
  networks: Abrupt change from second- to first-order transitions}}.
\newblock \emph{\bibinfo{journal}{Phys. Rev. E}} \textbf{\bibinfo{volume}{84}},
  \bibinfo{pages}{066116} (\bibinfo{year}{2011}).
\newblock \urlprefix\url{http://link.aps.org/doi/10.1103/PhysRevE.84.066116}.

\bibitem{bashan-pre2011}
\bibinfo{author}{Bashan, A.}, \bibinfo{author}{Parshani, R.} \&
  \bibinfo{author}{Havlin, S.}
\newblock \bibinfo{title}{{Percolation in networks composed of connectivity and
  dependency links}}.
\newblock \emph{\bibinfo{journal}{Phys. Rev. E}} \textbf{\bibinfo{volume}{83}},
  \bibinfo{pages}{051127} (\bibinfo{year}{2011}).
\newblock \urlprefix\url{http://link.aps.org/doi/10.1103/PhysRevE.83.051127}.

\bibitem{parshani-pnas2011}
\bibinfo{author}{Parshani, R.}, \bibinfo{author}{Buldyrev, S.~V.} \&
  \bibinfo{author}{Havlin, S.}
\newblock \bibinfo{title}{{Critical effect of dependency groups on the function
  of networks}}.
\newblock \emph{\bibinfo{journal}{Proceedings of the National Academy of
  Sciences}} \textbf{\bibinfo{volume}{108}}, \bibinfo{pages}{1007--1010}
  (\bibinfo{year}{2011}).
\newblock \urlprefix\url{http://www.pnas.org/content/108/3/1007.abstract}.

\bibitem{bashan-jsp2011}
\bibinfo{author}{Bashan, A.} \& \bibinfo{author}{Havlin, S.}
\newblock \bibinfo{title}{{The Combined Effect of Connectivity and Dependency
  Links on Percolation of Networks}}.
\newblock \emph{\bibinfo{journal}{Journal of Statistical Physics}}
  \textbf{\bibinfo{volume}{145}}, \bibinfo{pages}{686--695}
  (\bibinfo{year}{2011}).
\newblock \urlprefix\url{http://dx.doi.org/10.1007/s10955-011-0333-5}.

\bibitem{PhysRevE.85.066134}
\bibinfo{author}{Gao, J.}, \bibinfo{author}{Buldyrev, S.~V.},
  \bibinfo{author}{Havlin, S.} \& \bibinfo{author}{Stanley, H.~E.}
\newblock \bibinfo{title}{{Robustness of a network formed by $n$ interdependent
  networks with a one-to-one correspondence of dependent nodes}}.
\newblock \emph{\bibinfo{journal}{Phys. Rev. E}} \textbf{\bibinfo{volume}{85}},
  \bibinfo{pages}{066134} (\bibinfo{year}{2012}).
\newblock \urlprefix\url{http://link.aps.org/doi/10.1103/PhysRevE.85.066134}.

\bibitem{buldyrev-nature2010}
\bibinfo{author}{Buldyrev, S.~V.}, \bibinfo{author}{Parshani, R.},
  \bibinfo{author}{Paul, G.}, \bibinfo{author}{Stanley, H.~E.} \&
  \bibinfo{author}{Havlin, S.}
\newblock \bibinfo{title}{{Catastrophic cascade of failures in interdependent
  networks}}.
\newblock \emph{\bibinfo{journal}{Nature}} \textbf{\bibinfo{volume}{464}},
  \bibinfo{pages}{1025--1028} (\bibinfo{year}{2010}).
\newblock \urlprefix\url{http://dx.doi.org/10.1038/nature08932}.

\bibitem{gao-naturephysics2012}
\bibinfo{author}{Gao, J.}, \bibinfo{author}{Buldyrev, S.~V.},
  \bibinfo{author}{Stanley, H.~E.} \& \bibinfo{author}{Havlin, S.}
\newblock \bibinfo{title}{{Networks formed from interdependent networks}}.
\newblock \emph{\bibinfo{journal}{Nature Physics}}
  \textbf{\bibinfo{volume}{8}}, \bibinfo{pages}{40--48} (\bibinfo{year}{2012}).
\newblock \urlprefix\url{http://dx.doi.org/10.1038/nphys2180}.

\bibitem{vespignani-nature2010}
\bibinfo{author}{Vespignani, A.}
\newblock \bibinfo{title}{{Complex networks: The fragility of
  interdependency}}.
\newblock \emph{\bibinfo{journal}{Nature}} \textbf{\bibinfo{volume}{464}},
  \bibinfo{pages}{984--985} (\bibinfo{year}{2010}).
\newblock \urlprefix\url{http://dx.doi.org/10.1038/464984a}.

\bibitem{dong-preprint2012}
\bibinfo{author}{{Zhou}, D.}, \bibinfo{author}{{Bashan}, A.},
  \bibinfo{author}{{Berezin}, Y.}, \bibinfo{author}{{Cohen}, R.} \&
  \bibinfo{author}{{Havlin}, S.}
\newblock \bibinfo{title}{{On the Dynamics of Cascading Failures in
  Interdependent Networks}}.
\newblock \emph{\bibinfo{journal}{ArXiv e-prints}}  (\bibinfo{year}{2012}).
\newblock \eprint{1211.2330}.

\bibitem{bashan-naturephysics2013}
\bibinfo{author}{Bashan, A.}, \bibinfo{author}{Berezin, Y.},
  \bibinfo{author}{Buldyrev, S.~V.} \& \bibinfo{author}{Havlin, S.}
\newblock \bibinfo{title}{{The extreme vulnerability of interdependent
  spatially embedded networks}}.
\newblock \emph{\bibinfo{journal}{Nature Physics}}
  \textbf{\bibinfo{volume}{9}}, \bibinfo{pages}{667--672}
  (\bibinfo{year}{2013}).
\newblock \urlprefix\url{http://dx.doi.org/10.1038/nphys2727}.

\bibitem{rinaldi-ieee2001}
\bibinfo{author}{Rinaldi, S.}, \bibinfo{author}{Peerenboom, J.} \&
  \bibinfo{author}{Kelly, T.}
\newblock \bibinfo{title}{{Identifying, understanding, and analyzing critical
  infrastructure interdependencies}}.
\newblock \emph{\bibinfo{journal}{Control Systems, IEEE}}
  \textbf{\bibinfo{volume}{21}}, \bibinfo{pages}{11--25}
  (\bibinfo{year}{2001}).

\bibitem{peerenboom-proceedings2001}
\bibinfo{author}{Peerenboom, J.}, \bibinfo{author}{Fischer, R.} \&
  \bibinfo{author}{Whitfield, R.}
\newblock \bibinfo{title}{{Recovering from disruptions of interdependent
  critical infrastructures}}.
\newblock In \emph{\bibinfo{booktitle}{{Proc. CRIS/DRM/IIIT/NSF Workshop
  Mitigat. Vulnerab. Crit. Infrastruct. Catastr. Failures}}}
  (\bibinfo{year}{2001}).

\bibitem{radicchi-naturephysics2013}
\bibinfo{author}{Radicchi, F.} \& \bibinfo{author}{Arenas, A.}
\newblock \bibinfo{title}{{Abrupt transition in the structural formation of
  interconnected networks}}.
\newblock \emph{\bibinfo{journal}{Nature Physics}}
  \textbf{\bibinfo{volume}{9}}, \bibinfo{pages}{717--720}
  (\bibinfo{year}{2013}).

\bibitem{son-epl2012}
\bibinfo{author}{Son, S.-W.}, \bibinfo{author}{Bizhani, G.},
  \bibinfo{author}{Christensen, C.}, \bibinfo{author}{Grassberger, P.} \&
  \bibinfo{author}{Paczuski, M.}
\newblock \bibinfo{title}{{Percolation theory on interdependent networks based
  on epidemic spreading}}.
\newblock \emph{\bibinfo{journal}{EPL (Europhysics Letters)}}
  \textbf{\bibinfo{volume}{97}}, \bibinfo{pages}{16006} (\bibinfo{year}{2012}).
\newblock \urlprefix\url{http://stacks.iop.org/0295-5075/97/i=1/a=16006}.

\bibitem{serrano-pre2012}
\bibinfo{author}{Saumell-Mendiola, A.}, \bibinfo{author}{Serrano, M.~{\'A}.} \&
  \bibinfo{author}{Bogu{\~n}{\'a}, M.}
\newblock \bibinfo{title}{{Epidemic spreading on interconnected networks}}.
\newblock \emph{\bibinfo{journal}{Phys. Rev. E}} \textbf{\bibinfo{volume}{86}},
  \bibinfo{pages}{026106} (\bibinfo{year}{2012}).
\newblock \urlprefix\url{http://link.aps.org/doi/10.1103/PhysRevE.86.026106}.

\bibitem{morris-prl2012}
\bibinfo{author}{Morris, R.~G.} \& \bibinfo{author}{Barthelemy, M.}
\newblock \bibinfo{title}{{Transport on Coupled Spatial Networks}}.
\newblock \emph{\bibinfo{journal}{Phys. Rev. Lett.}}
  \textbf{\bibinfo{volume}{109}}, \bibinfo{pages}{128703}
  (\bibinfo{year}{2012}).
\newblock
  \urlprefix\url{http://link.aps.org/doi/10.1103/PhysRevLett.109.128703}.

\bibitem{zhao-jstatmech2013}
\bibinfo{author}{Zhao, K.} \& \bibinfo{author}{Bianconi, G.}
\newblock \bibinfo{title}{{Percolation on interacting, antagonistic networks}}.
\newblock \emph{\bibinfo{journal}{Journal of Statistical Mechanics: Theory and
  Experiment}} \textbf{\bibinfo{volume}{2013}}, \bibinfo{pages}{P05005}
  (\bibinfo{year}{2013}).
\newblock \urlprefix\url{http://stacks.iop.org/1742-5468/2013/i=05/a=P05005}.

\bibitem{donges-epjb2011}
\bibinfo{author}{Donges, J.}, \bibinfo{author}{Schultz, H.},
  \bibinfo{author}{Marwan, N.}, \bibinfo{author}{Zou, Y.} \&
  \bibinfo{author}{Kurths, J.}
\newblock \bibinfo{title}{Investigating the topology of interacting networks}.
\newblock \emph{\bibinfo{journal}{The European Physical Journal B}}
  \textbf{\bibinfo{volume}{84}}, \bibinfo{pages}{635--651}
  (\bibinfo{year}{2011}).
\newblock \urlprefix\url{http://dx.doi.org/10.1140/epjb/e2011-10795-8}.

\bibitem{gao-general-net}
\bibinfo{author}{Gao, J.}, \bibinfo{author}{Buldyrev, S.~V.},
  \bibinfo{author}{Stanley, H.~E.}, \bibinfo{author}{Xu, X.} \&
  \bibinfo{author}{Havlin, S.}
\newblock \bibinfo{title}{Percolation of a general network of networks}.
\newblock \emph{\bibinfo{journal}{Phys. Rev. E}} \textbf{\bibinfo{volume}{88}},
  \bibinfo{pages}{062816} (\bibinfo{year}{2013}).
\newblock \urlprefix\url{http://link.aps.org/doi/10.1103/PhysRevE.88.062816}.

\bibitem{wei-prl2012}
\bibinfo{author}{Li, W.}, \bibinfo{author}{Bashan, A.},
  \bibinfo{author}{Buldyrev, S.~V.}, \bibinfo{author}{Stanley, H.~E.} \&
  \bibinfo{author}{Havlin, S.}
\newblock \bibinfo{title}{{Cascading Failures in Interdependent Lattice
  Networks: The Critical Role of the Length of Dependency Links}}.
\newblock \emph{\bibinfo{journal}{Phys. Rev. Lett.}}
  \textbf{\bibinfo{volume}{108}}, \bibinfo{pages}{228702}
  (\bibinfo{year}{2012}).
\newblock
  \urlprefix\url{http://link.aps.org/doi/10.1103/PhysRevLett.108.228702}.

\bibitem{agarwal-milcom2010}
\bibinfo{author}{Agarwal, P.~K.} \emph{et~al.}
\newblock \bibinfo{title}{Network vulnerability to single, multiple, and
  probabilistic physical attacks}.
\newblock In \emph{\bibinfo{booktitle}{MILITARY COMMUNICATIONS CONFERENCE, 2010
  - MILCOM 2010}}, \bibinfo{pages}{1824--1829} (\bibinfo{year}{2010}).

\bibitem{wang-proceedings2008}
\bibinfo{author}{Wang, Z.}, \bibinfo{author}{Thomas, R.~J.} \&
  \bibinfo{author}{Scaglione, A.}
\newblock \emph{\bibinfo{title}{{Generating Random Topology Power Grids}}},
  \bibinfo{pages}{183--183} (\bibinfo{publisher}{Institute of Electrical and
  Electronics Engineers}, \bibinfo{year}{2008}).
\newblock \urlprefix\url{http://dx.doi.org/10.1109/HICSS.2008.182}.

\bibitem{albert-pre2004}
\bibinfo{author}{Albert, R.}, \bibinfo{author}{Albert, I.} \&
  \bibinfo{author}{Nakarado, G.~L.}
\newblock \bibinfo{title}{{Structural vulnerability of the North American power
  grid}}.
\newblock \emph{\bibinfo{journal}{Phys. Rev. E}} \textbf{\bibinfo{volume}{69}},
  \bibinfo{pages}{025103} (\bibinfo{year}{2004}).
\newblock \urlprefix\url{http://link.aps.org/doi/10.1103/PhysRevE.69.025103}.

\bibitem{kinney-epjb2005}
\bibinfo{author}{Kinney, R.}, \bibinfo{author}{Crucitti, P.},
  \bibinfo{author}{Albert, R.} \& \bibinfo{author}{Latora, V.}
\newblock \bibinfo{title}{{Modeling cascading failures in the North American
  power grid}}.
\newblock \emph{\bibinfo{journal}{The European Physical Journal B - Condensed
  Matter and Complex Systems}} \textbf{\bibinfo{volume}{46}},
  \bibinfo{pages}{101--107} (\bibinfo{year}{2005}).
\newblock \urlprefix\url{http://dx.doi.org/10.1140/epjb/e2005-00237-9}.

\bibitem{chassin-physa2005}
\bibinfo{author}{Chassin, D.~P.} \& \bibinfo{author}{Posse, C.}
\newblock \bibinfo{title}{{Evaluating North American electric grid reliability
  using the Barab{\'a}si--Albert network model}}.
\newblock \emph{\bibinfo{journal}{Physica A: Statistical Mechanics and its
  Applications}} \textbf{\bibinfo{volume}{355}}, \bibinfo{pages}{667--677}
  (\bibinfo{year}{2005}).
\newblock
  \urlprefix\url{http://www.sciencedirect.com/science/article/pii/S0378437105002311}.

\bibitem{hines-chaos2010}
\bibinfo{author}{Hines, P.}, \bibinfo{author}{Cotilla-Sanchez, E.} \&
  \bibinfo{author}{Blumsack, S.}
\newblock \bibinfo{title}{Do topological models provide good information about
  electricity infrastructure vulnerability?}
\newblock \emph{\bibinfo{journal}{Chaos: An Interdisciplinary Journal of
  Nonlinear Science}} \textbf{\bibinfo{volume}{20}}, \bibinfo{pages}{033122}
  (\bibinfo{year}{2010}).
\newblock \urlprefix\url{http://link.aip.org/link/?CHA/20/033122/1}.

\bibitem{carreras-ieee2004}
\bibinfo{author}{Carreras, B.}, \bibinfo{author}{Newman, D.},
  \bibinfo{author}{Dobson, I.} \& \bibinfo{author}{Poole, A.}
\newblock \bibinfo{title}{{Evidence for self-organized criticality in a time
  series of electric power system blackouts}}.
\newblock \emph{\bibinfo{journal}{Circuits and Systems I: Regular Papers, IEEE
  Transactions on}} \textbf{\bibinfo{volume}{51}}, \bibinfo{pages}{1733--1740}
  (\bibinfo{year}{2004}).

\bibitem{barthelemy-physicsreports2011}
\bibinfo{author}{Barth{\'e}lemy, M.}
\newblock \bibinfo{title}{Spatial networks}.
\newblock \emph{\bibinfo{journal}{Physics Reports}}
  \textbf{\bibinfo{volume}{499}}, \bibinfo{pages}{1 -- 101}
  (\bibinfo{year}{2011}).

\bibitem{bunde1991fractals}
\bibinfo{author}{Bunde, A.} \& \bibinfo{author}{Havlin, S.}
\newblock \emph{\bibinfo{title}{{Fractals and disordered systems}}}
  (\bibinfo{publisher}{Springer-Verlag New York, Inc.}, \bibinfo{year}{1991}).

\bibitem{danziger_interdependent}
\bibinfo{author}{Danziger, M.~M.}, \bibinfo{author}{Bashan, A.},
  \bibinfo{author}{Berezin, Y.} \& \bibinfo{author}{Havlin, S.}
\newblock \bibinfo{title}{Interdependent spatially embedded networks: Dynamics
  at percolation threshold}.
\newblock In \emph{\bibinfo{booktitle}{Signal-Image Technology \&
  Internet-Based Systems (SITIS), 2013 International Conference on}},
  \bibinfo{pages}{619--625} (\bibinfo{organization}{IEEE},
  \bibinfo{year}{2013}).

\bibitem{zhou2012critical}
\bibinfo{author}{Zhou, D.}, \bibinfo{author}{Bashan, A.},
  \bibinfo{author}{Berezin, Y.}, \bibinfo{author}{Shnerb, N.} \&
  \bibinfo{author}{Havlin, S.}
\newblock \bibinfo{title}{Critical cascading failures in interdependent
  networks: Non mean-field behavior}.
\newblock \emph{\bibinfo{journal}{arXiv preprint arXiv:1211.2330}}
  (\bibinfo{year}{2012}).

\end{thebibliography}

\end{document}